\documentclass[fleqn,usenatbib]{mnras}

\usepackage{newtxtext,newtxmath}

\usepackage[T1]{fontenc}

\DeclareRobustCommand{\VAN}[3]{#2}
\let\VANthebibliography\thebibliography
\def\thebibliography{\DeclareRobustCommand{\VAN}[3]{##3}\VANthebibliography}

\usepackage{graphicx}	
\usepackage{amsmath}	
\usepackage{orcidlink}
\usepackage{bm} 
\usepackage{multirow}
\hypersetup{linkcolor=red,citecolor=orange,filecolor=cyan,urlcolor=magenta}

\newcommand{\gist}{\texttt{GIST}}
\newcommand{\profound}{\texttt{ProFound}}
\newcommand{\esc}{erg s$^{-1}$ cm$^{-2}$}
\newcommand{\zspec}{$z_\mathrm{spec}$}
\newcommand{\EBVgas}{$E(B-V)_{\mathrm{gas}}$}

\newcommand{\EBVratioavg}{$\langle E(B-V)_{\mathrm{star}}\rangle/\langle E(B-V)_{\mathrm{gas}}\rangle$}
\newcommand{\halpha}{$\mathrm{H}\alpha$}
\newcommand{\hbeta}{$\mathrm{H}\beta$}

\title[MAGPI Emission Line Products DR]{The MAGPI Survey\thanks{Based on observations obtained at the Very Large Telescope (VLT) of the European Southern Observatory (ESO), Paranal, Chile (ESO program ID 1104.B-0536)}: Emission Line Products Data Release and the Role of Spectroscopic Aperture Covering Fraction on the Balmer Decrement-Stellar Mass Relation}

\author[A. J. Battisti et al.]{A. J. Battisti\orcidlink{0000-0003-4569-2285}$^{1,2,3}$\thanks{Email:andrew.battisti@uwa.edu.au},
E. G. M. Muller\orcidlink{0000-0001-5621-1577}$^{2,3}$,
E. Wisnioski\orcidlink{0000-0003-1657-7878}$^{2,3}$,
J. T. Mendel\orcidlink{0000-0002-6327-9147}$^{2,3}$,
C. Foster\orcidlink{0000-0003-0247-1204}$^{4,3}$,
\newauthor 
C. D. P. Lagos\orcidlink{0000-0003-3021-8564}$^{1,3}$,
K. E. Harborne\orcidlink{0000-0002-2043-7985}$^{5,6,3}$,
I. U. Aalia\orcidlink{0009-0005-0489-6232}$^{7,8}$,
S. Barsanti\orcidlink{0000-0002-9332-5386}$^{9}$,
I. Breda\orcidlink{0000-0002-3764-5780}$^{10}$,
D. Calzetti\orcidlink{0000-0002-5189-8004}$^{11}$,
\newauthor 
S. M. Croom\orcidlink{0000-0003-2880-9197}$^{9}$,
P. K. Das\orcidlink{0000-0002-4326-8598}$^{12}$,
A. Ferré-Mateu\orcidlink{0000-0002-6411-220X}$^{13,14,15}$,
T. Gao\orcidlink{0000-0002-1158-6372}$^{2,3}$,
E. Gjergo\orcidlink{0000-0002-7440-1080}$^{16,17}$,
K. Grasha\orcidlink{0000-0002-3247-5321}$^{2,3}$\thanks{ARC DECRA Fellow}, %
\newauthor 
Y. Mai\orcidlink{0000-0003-3514-6280}$^{18,2,8}$,
A. Mailvaganam\orcidlink{0009-0003-1221-1630}$^{7,8,2}$,
T. Mukherjee\orcidlink{0009-0004-7639-869X}$^{7,8}$,
M. Mun\orcidlink{0000-0002-3706-9955}$^{19}$,
R.-S. Remus\orcidlink{0009-0008-9260-7278}$^{26,15}$,
\newauthor 
G. Sharma\orcidlink{0000-0002-6070-2851}$^{20,21,22,23,24}$,
S. M. Sweet\orcidlink{0000-0002-1576-2505}$^{12,2}$,
S. Thater\orcidlink{0000-0003-1820-2041}$^{25}$,
L. M. Valenzuela\orcidlink{0000-0002-7972-9675}$^{26}$,
J. van de Sande\orcidlink{0000-0003-2552-0021}$^{4}$,
\newauthor 
G. van de Ven\orcidlink{0000-0003-4546-7731}$^{25}$,
T. Zafar\orcidlink{0000-0003-3935-7018}$^{7,8}$,
B. Ziegler\orcidlink{0000-0003-2856-1080}$^{25}$
\\
Affiliations are listed at the end of the paper
}

\date{Accepted XXX. Received YYY; in original form ZZZ}

\pubyear{\the\year{}}

\begin{document}
\label{firstpage}
\pagerange{\pageref{firstpage}--\pageref{lastpage}}
\maketitle

\begin{abstract}
The Middle Ages Galaxy Properties with Integral field spectroscopy (MAGPI) survey is a Large Program on the European Southern Observatory Very Large Telescope using the MUSE instrument. 
This paper presents the data release for the MAGPI emission line products and includes emission line maps for 836 galaxies at $0.05\leq z_\mathrm{spec}\leq 0.424$ (\halpha-window) and aperture-based emission line measurements for 2,607 galaxies at $0.05\leq z_\mathrm{spec}\leq 1.50$ (upper bound is [OII] cut-off), both based on the \gist\ software, for all 56 MAGPI fields. 
We use these data to examine dust attenuation, which represents a major source of uncertainty in the derived properties of galaxies that are critical to constrain models of galaxy evolution. We examine the role that the spectroscopic aperture covering fraction ($f_c$) has on the relationship between the Balmer decrement ($\mathrm{BD}=F(\mathrm{H}\alpha)/F(\mathrm{H}\beta)$; a common proxy for dust attenuation) and the total stellar mass ($M_\star$). Several studies have suggested that the BD-$M_\star$ relation may be redshift invariant; however, the compared surveys often have different $f_c$ due to their differing fibre or slit sizes that can cause systematic offsets. Our results indicate that $f_c$ has a significant impact on this relationship, due to galaxies having negative BD radial gradients, which are more negative for more massive galaxies at $z\lesssim0.4$. Comparing spectroscopic surveys with $\left<f_c\right> \gtrsim 0.5$, we find that the BD-$M_\star$ relation shows a preference for redshift evolution and is roughly consistent with the behaviour of UV stellar continuum attenuation redshift evolution ($A_\mathrm{FUV}$-$z$), with the average dust attenuation in galaxies peaking at $z\sim1.2$ and decreasing at lower and higher redshifts. 
\end{abstract}




\section{Introduction}\label{intro}

Integral Field Spectroscopic (IFS) surveys are a powerful means to study galaxy evolution by providing spatially resolved spectroscopic measurements of stars and ionised gas on a region-by-region basis within galaxies \citep{sanchez20}. These measurements can constrain key physical properties and kinematics of the stellar population (e.g., age, stellar metallicity) and the interstellar medium (e.g., nebular dust attenuation, gas-phase metallicity).
The largest optical IFS programs that targeted \textit{both} star-forming and elliptical galaxies at low redshifts ($z\lesssim0.15$) have been the extended-CALIFA \citep[$\sim$900 galaxies;][]{sanchez12, sanchez23}, MaNGA \citep[$\sim$10,000 galaxies;][]{bundy15, abdurrouf22}, SAMI \citep[$\sim$3,000 galaxies;][]{croom12, croom21}, and AMUSING++ \citep[MUSE compilation, $\sim$600 galaxies;][]{lopez-coba20} surveys; noting also the ongoing Hector survey \citep[$\sim$15,000 galaxies;][]{bryant24, oh25}. Large sample sizes ($N\gtrsim200$) are important for robust statistical analyses to disentangle confounding factors, such as internal and external feedback processes \citep{cortese21}, that influence galaxy evolution. These large surveys have resulted in major gains in our understanding of the roles that various internal and external processes have on the physical properties, kinematics, and morphology of galaxies \citep[see][and references therein]{sanchez20}. However, a shortcoming of these surveys is that they are limited to galaxies at $z\lesssim0.15$ where galaxies can be well-resolved without the use of adaptive optics on ground-based facilities. This, in turn, limits the ability to characterise the extent to which these internal and external physical processes may evolve over cosmic time. For example, observations \citep[e.g.,][]{darvish16, bezanson18} and simulations \citep[e.g.,][]{schaye15, teklu15, pillepich18} suggest that environmental processes begin to dominate galaxy evolution over internal processes at $z\lesssim1$; however, it is not clear exactly at what redshift this change occurs. Simulations show large differences in predictions of galaxy rotational support ($v/\sigma$) at $z\sim0.3$ \citep[e.g.,][see their Figure~1]{foster21} that can be tested with IFS observations at these redshifts.

The Middle Ages Galaxy Properties with Integral field spectroscopy (MAGPI) survey \citep{foster21}, a 340~hr VLT/MUSE Large Program, is a step toward examining the structural transformation of hundreds of galaxies at a lookback time of 3–4 Gyr ($0.26\leq z\leq 0.42$). This is achieved through the use of adaptive optics on the 8.2m VLT UT4 and represents the practical limit in terms of sensitivity and spatial resolution that can be achieved in IFS surveys with current ground-based facilities. 
At higher redshift ($z\gtrsim0.5$), considerable IFS samples have been obtained (see \citealt{forsterSchreiber&wuyts20} and references therein) with new samples emerging from the James Webb Space Telescope (JWST; e.g., GA-NIFS\footnote{https://ga-nifs.github.io}; MSA-3D, \citealt{barisic25}). 
However, these samples are comparatively smaller and limited by spatial resolution, sensitivity, and/or atmospheric features (ground-based) that limit simultaneous wavelength coverage, which is ideal for studies of the dust content and metallicity of the ISM.

Deriving accurate physical properties of galaxies from extragalactic photometric surveys covering a wide range in redshift is critical to improving our understanding of galaxy evolution across cosmic time. However, a key limitation that reduces this accuracy is uncertainties in corrections for dust attenuation of light spanning ultraviolet to near-infrared wavelengths. The two primary ways in which these corrections are typically made for individual galaxies are to use dust-sensitive emission line diagnostics (e.g., Balmer decrement, $\mathrm{BD}=F(\mathrm{H}\alpha)/F(\mathrm{H}\beta)$) and/or rest-frame infrared (IR) dust emission (assuming energy balance with dust absorption). However, neither spectroscopy of emission lines nor IR data are typically available for most photometric galaxy surveys, especially at $z\gtrsim1$, and as a result simple average scaling relations between dust attenuation and galaxy properties that are only weakly dependent on dust (e.g., total stellar mass) are often adopted instead to correct for dust. 

One popular scaling relation is the correlation between the Balmer decrement and total stellar mass of galaxies (hereafter BD-$M_\star$ relation), which exists at low-$z$ \citep[$z\lesssim0.3$; e.g.,][]{garn&best10, zahid13, battisti16}, cosmic noon \citep[$1\lesssim z\lesssim2$; e.g.,][]{battisti22, shapley22, matharu23, liu25}, and high-$z$ \citep[$z\gtrsim3$; e.g.,][]{shapley23, woodrum25, faisst26}.
This relation serves as a rough way to make \textit{average} dust corrections of emission lines\footnote{The reddening of the stellar continuum is typically lower by roughly a factor of two relative to the reddening of ionised gas (traced by the BD) in local galaxies, but this has considerable scatter \citep[\EBVratioavg$\ \sim0.5$; e.g.,][]{calzetti94, kreckel13, battisti16, zahid17, emsellem22, battisti25a}. Therefore, it is not recommended to use the BD-$M_\star$ relationship to make corrections of stellar continuum-related quantities.} for moderate to large samples of galaxies \citep[e.g., see][]{battisti24}. This is relevant for application to large grism samples from \textit{Euclid} and \textit{Roman} when limited emission lines are available (e.g., single or no hydrogen lines to determine the BD). 

An unresolved aspect of the BD-$M_\star$ relation is whether it evolves with redshift, which is difficult to establish because the commonly adopted local relation from \citet{garn&best10} is based on SDSS-III \citep{abazajian09}, which used 3\arcsec\ diameter fibres for spectroscopy that only enclose $\sim$25\% of the stellar mass on average \citep[e.g.,][]{zahid17}. This influences the BD-$M_\star$ relation because the BD is measured only within the fibre region whereas the stellar mass from the whole galaxy is typically adopted. Since galaxies show radial gradients in their dust attenuation (dust content) with larger attenuation (more dust) residing in the inner regions than the outer regions \citep[e.g.,][]{greener20}, it is possible that the normalisation and shape of this relation may change if larger apertures are adopted that cover the majority of each galaxy. 
IFS surveys like SAMI ($z\sim0$) and MAGPI ($z\sim0.3$) provide a straightforward manner to examine the impact that spectroscopic aperture covering fractions have on the BD-$M_\star$ relation at low-$z$ because their fields-of-view (FoVs) cover the majority of optical galaxy extents.   

This paper presents the release of emission line data products developed by the MAGPI survey team. These products include emission line maps for 836 galaxies at $0.05\leq z_\mathrm{spec}\leq 0.424$ (\halpha\ window) and aperture-based emission line measurements for 2,607 galaxies at $0.05\leq z_\mathrm{spec}\leq 1.50$. We provide a basic overview of the sample properties and demonstrate its science application to examine the BD gradients of galaxies and the BD-$M_\star$ relation and its evolution.
At $2.9 < z < 6.7$, the Lyman-$\alpha$ feature is covered by MUSE, allowing detailed studies of ISM and circumgalactic medium gas kinematics in distant galaxies \citep{mukherjee24, mukherjee26}. A separate data release of Lyman-$\alpha$ emitters will be presented in Mukherjee et al. (in prep.). Other upcoming MAGPI data releases include reduced MUSE datacubes (Mendel et al. in prep.), stellar absorption line data products (Poci et al. in prep.), and theory data products (Harborne et al. in prep.).

The paper is organised as follows. Section~\ref{data} describes the observational data of the MAGPI and SAMI surveys (latter used as $z\sim0$ benchmark). Section~\ref{EL_pipeline} describes the MAGPI emission line pipeline. Section~\ref{EL_products} presents the MAGPI emission line data products. Section~\ref{sample_selection} describes the selection criteria used for the sample in our results. Section~\ref{results} presents our results that include an overview of the properties of the MAGPI emission line sample (Section~\ref{sample_overview}) and an application of the data products to study BD radial gradients (Section~\ref{BD_radius_compare}), the BD-$M_\star$ relation (Section~\ref{BD_mstar_compare}), and the BD correlation strength with various galaxy properties (Section~\ref{BD_correlation_compare}). Section~\ref{discussion} compares our BD-$M_\star$ relations with the literature and discusses its redshift evolution. Section~\ref{conclusion} summarises our main conclusions. Throughout this work, we adopt a $\Lambda$-CDM cosmological model, with $H_0=70$~km/s/Mpc, $\Omega_M=0.3$, and $\Omega_{\Lambda}=0.7$. All magnitudes are in the AB magnitude system \citep{oke&gunn83}.

\section{Data}\label{data}

\subsection{MAGPI Observations and Data Reduction}\label{MAGPI_data}
\begin{figure*}
	\includegraphics[width=\textwidth]{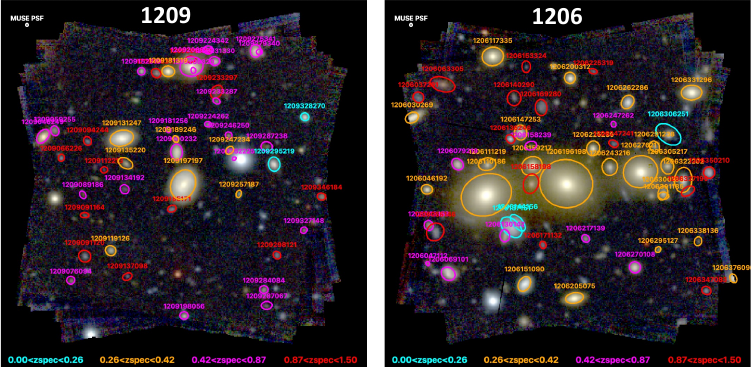}
 \vspace{-5mm}
    \caption{RGB images created from $g_\mathrm{mod}$ (blue), $r$ (green), and $i$ (red) pseudo-filters for two MAGPI MUSE datacubes (IDs shown in upper-middle of each panel). Galaxies with confident \zspec\ (\texttt{QOP}=3 or 4) are indicated by coloured ellipses and IDs, with the colours corresponding to different redshift ranges (cyan: $0.00\leq z_\mathrm{spec} \leq 0.26$; orange: $0.26\leq z_\mathrm{spec} \leq 0.42$; magenta: $0.42\leq z_\mathrm{spec} \leq 0.87$; red: $0.87\leq z_\mathrm{spec} \leq 1.50$) where key emission lines shift in/out of MUSE coverage (see Table~\ref{tab:emission_lines}). The size of the ellipses correspond to the radii enclosing 90\% of the $i$-band flux ($r_{90,i}$). The MAGPI fields were chosen with the primary goal to study hundreds of spatially-resolved galaxies at $z\sim 0.3$ (orange) that span a range of environments, parameterised through the group halo mass ($M_\mathrm{halo}$), with the 1206 field (right) representing a denser environment than the 1209 field (left).  It is evident that these fields include many serendipitous sources (cyan, magenta, red) that can be used for additional science. This data release includes emission line measurements for all galaxies with $z_\mathrm{spec} \leq 1.50$. Labelled images and SAOImageDS9 region files (\texttt{*.reg}) providing the ellipses and IDs for each field are available in the online version as a figureset.
    \label{fig:MAGPI_environment_examples}}
\end{figure*}

\begin{figure}
	\includegraphics[width=0.5\textwidth]{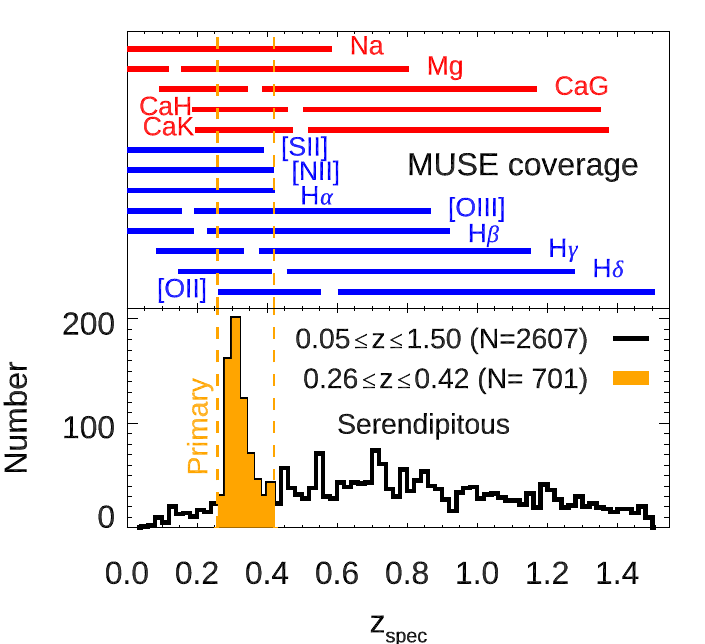}
 \vspace{-5mm}
    \caption{\textit{Top:} MUSE coverage for optical emission lines (blue; listed in Table~\ref{tab:emission_lines}) and absorption lines (red). The small gap in coverage for each line is due to a notch filter to block the GLAO sodium laser. \textit{Bottom:} Redshift histogram of galaxies with secure redshifts based on two or more emission and/or absorption features (\texttt{QOP}=3 or 4) in all 56 MAGPI fields. The MAGPI primary sample spans $0.26\leq z_\mathrm{spec} \leq 0.42$ (orange), corresponding to the window with complete coverage of emission lines from [OII] to [NII] to characterise the interstellar medium, as well as key optical absorption features to characterise stellar populations. These fields also include thousands of serendipitous sources that can be used for additional science. 
    \label{fig:MAGPI_zhist}}
\end{figure}

The MAGPI survey \citep[][Program ID: 1104.B-0536]{foster21, mendel24} is a VLT/MUSE Large Program (340~hr) using ground-layer adaptive optics (GLAO), which provides excellent spatial resolution ($i$-band PSF FWHM$\sim$0.55\arcsec). The MUSE instrument provides spatially-resolved spectroscopy over 470-935~nm ($R\sim1800-3600$) over an area of 1~arcmin$^2$. However, there is a small gap in spectral coverage from 580-597~nm due to a notch filter to block the GLAO sodium laser. 
MAGPI consists of 56 independent fields (covering 65~arcmin$^2$ in total\footnote{A coadded MAGPI-MUSE datacube covers $\sim$1.17~arcmin$^2$ \citep[][]{foster21}, which is larger than a single MUSE datacube, due to the observing strategy (dither and rotation).}) that were chosen to span a large dynamic range in galaxy environmental density (isolated/group/cluster; Figure~\ref{fig:MAGPI_environment_examples}; see \citealt{foster21}) at $0.26\leq z\leq 0.42$ (3–4 Gyr lookback time). This redshift range corresponds to the window where all strong rest-frame optical emission and absorption lines are covered by the MUSE instrument (see Figure~\ref{fig:MAGPI_zhist}) to probe ionised gas and stellar properties.
All 56 fields have been observed and are included in this data release. Five $z\sim0.3$ cluster fields with archival MUSE data \citep[ABELL0370, ABELL2744, MACS0257, MACS0416, MACS0940;][]{richard21}, which sample the densest environments, are also included in the survey (reprocessed through MAGPI pipeline; see Mendel et al. in prep.), but emission line products for those cluster fields will be included as a separate data release.

The MAGPI fields lie within the GAMA survey \citep[G12, G15, and G23 fields;][]{driver11, driver22} and have moderate depth optical--near-IR photometric data from the Kilo-Degree Survey \citep[KiDS, $ugri$ filters;][]{kuijken19} and VISTA Kilo-degree INfrared Galaxy \citep[VIKING, $ZYJHK_s$ filters;][]{wright19} surveys. These fields also have far-IR data from the Herschel-ATLAS survey \citep{eales10}, but these are too shallow to detect the majority of sources at $z\gtrsim0.2$ in the MAGPI fields.
The MAGPI fields in G12 and G15 also have imaging from Hyper Suprime-Cam Subaru Strategic Program \citep[HSC-SSP, $grizy$ filters;][]{aihara19}.

The details of the MAGPI MUSE data reduction are summarised in \cite{foster21} and will be described in detail in Mendel et al. (in prep). In brief, the data are first processed using the \texttt{PYMUSEPIPE}\footnote{\url{https://github.com/emsellem/pymusepipe}} software \citep{emsellem22}, a Python wrapper for the MUSE reduction pipeline \citep{weilbacher20}. \texttt{PYMUSEPIPE} performs wavelength calibration and measures the instrumental line-spread function (LSF). The illumination correction and sky subtraction are performed using the \texttt{CUBEFIX} \citep{cantalupo19} and \texttt{ZAP} \citep{soto16} software, respectively. Next, source detection, segmentation maps, and structural parameters (e.g., sizes, position angle, effective radii) are obtained by running \profound\footnote{\url{https://github.com/asgr/ProFound}} \citep{robotham18} on the post-processed `white-light' datacubes. \profound\ identifies objects that have a signal above $3\times\mathrm{RMS}_\mathrm{sky}$. The initial segmentation mask was manually adjusted to join mistakenly split segments or remove
visibly spurious detections, similar to the method described in \citet{bellstedt20a}. 
Despite best efforts to manually adjust segments, we caution that care should be taken for sources that spatially overlap because blended light is not directly accounted for in the spectral modelling in this data release. Users are encouraged to consult the \texttt{REDUCED\_CHI2} map products (see Section~\ref{EL_products}) to identify regions where blending significantly degrades the fit quality.
Additional continuum-faint, emission-line sources are found using custom software with segments added to the full segmentation map, and will be described in Mukherjee et al. (in prep). 

The MAGPI source naming convention is \texttt{MAGPI\#\#\#\#XXXYYY}, where \texttt{\#\#\#\#} is the field ID (e.g., `1201') and \texttt{XXX} and \texttt{YYY} are the x- and y-pixel positions of the brightest spaxel in $r$-band for a source in the full MAGPI-MUSE datacube. 
The full MAGPI-MUSE datacubes are 396$\times$396 spaxels in size \citep[covering $\sim$1.17~arcmin$^2$, excluding null areas;][]{foster21}. 
The first two digits of the field ID denote the GAMA region (G12, G15, or G23) and the last two are the field number within that region (starting at `01'). There are 9 MAGPI fields in the G12 region (MAGPI1201-MAGPI1209), 35 in the G15 region (MAGPI1501-MAGPI1535), and 12 in the G23 region (MAGPI2301-MAGPI2312).
The MUSE pointings were typically centred on group central galaxies, as defined in the GAMA group catalogue \citep{robotham11}, which is usually the brightest (often most massive) galaxy in each group. As a result, a large number of the group centrals are close to \texttt{XXXYYY}=197197 (e.g., \texttt{MAGPI1202197197}). 

To set the size of datacubes for individual galaxies (hereafter referred to as `minicubes'), the initial \profound\ segmentation map (referred to as the `undilated' mask), is extended outward until it reaches the sky noise level (referred to as the `dilated' mask). Then a rectangular region encompassing the dilated segmentation map is extracted with \texttt{MPDAF}\footnote{\url{https://github.com/musevlt/mpdaf}}. 
In addition to these minicubes, we also measure a series of integrated and aperture-based spectra that serve as additional inputs into the emission line pipeline (Section~\ref{EL_pipeline}). The details of the various apertures (referred to as `\texttt{onedspec}' products) will be presented in Section~\ref{aperture_products}.

The spectroscopic redshift, \zspec, (barycentric) for each galaxy is determined using a 1" radius circular aperture spectrum centred on the galaxy and using the \texttt{MARZ} software\footnote{\url{https://github.com/Samreay/Marz}} \citep{hinton16}. More details on redshift assignments will be presented in Mendel et al. (in prep). For this data release, products are generated only for sources with confident redshift assignments, corresponding to a redshift quality flag \texttt{QOP}=3 or 4, which occurs for 2,607 galaxies. The redshift histogram of the sample in this data release is shown in Figure~\ref{fig:MAGPI_zhist} and examples are highlighted in Figure~\ref{fig:MAGPI_environment_examples}.

\subsection{SAMI Comparison Sample}\label{SAMI_data}

As a $z\sim0$ reference for our Balmer decrement results, we construct a comparison sample from the SAMI Galaxy Survey \citep{croom12, bryant15} to match the emission line analysis applied to the MAGPI data. SAMI is an optical IFS instrument on the AAT that used hexabundles (fibre bundles) covering an area of 15\arcsec\ in diameter. We adopt SAMI as a reference sample, rather than other low-redshift samples (e.g., CALIFA, MaNGA), because the stellar mass maps derived for MAGPI use the same methodology that was used in SAMI \citep[see Section~\ref{sample_overview};][]{taylor11, medling18} for a more self-consistent comparison\footnote{The integrated aperture BD-$M_\star$ relation presented in Section~\ref{BD_mstar_compare} for SAMI appears consistent with that from MaNGA when using the SDSS DR17 \citep{abdurrouf22} \texttt{DAPall} and \texttt{DRPall} data products, which provide integrated emission line and stellar mass measurements, respectively.}.

The final data release of the SAMI Galaxy Survey \citep[DR3;][]{croom21} comprises 3,068 galaxies drawn from the GAMA fields and eight cluster regions, spanning a redshift range of $0.004<z<0.095$ and a stellar mass range of $10^7 - 10^{12}\ \textrm{M}_\odot$. We restrict the comparison sample to the main SAMI survey and exclude galaxies from the SAMI cluster sample \citep{owers17}, as stellar mass surface density maps are not available for the latter. Furthermore, we do not explore the role of environment on BD values in this work and therefore excluding the cluster sample is also preferred for this reason. 

We use the following DR3 products \citep{croom21}:
(1) circular aperture line measurements from the 1-component catalogue (\texttt{sami\_dr3.EmissionLine1compDR3}) and 
(2) Recommended-component emission-line maps (\texttt{[ID]\_[letter]\_[line]\_default\_recom-comp.fits}), where \texttt{[letter]} (i.e., \texttt{A}, \texttt{B}) refers to the visit for a source (some sources have multiple observations). The best visit is indicated using the \texttt{ISBEST} flag, with the \texttt{CUBEIDPUB} column in the \texttt{sami\_dr3.CubeObs} catalogue indicating the visit (\texttt{[ID]\_[letter]}).
The emission line measurements are derived using the spectral fitting code \texttt{LZIFU} \citep{ho16}. 
We use stellar mass surface density maps from \citet{medling18} to derive aperture stellar mass values for each galaxy. Total stellar masses are taken from the GAMA input catalogue \citep[\texttt{sami\_dr3.InputCatGAMADR3},][]{bryant15}, which is based on photometry, and thus is representative of the total mass regardless of the covering fraction of the SAMI hexabundle.

\section{MAGPI Emission Line Products Pipeline}\label{EL_pipeline} 
\begin{figure*}
	\includegraphics[width=0.99\textwidth]{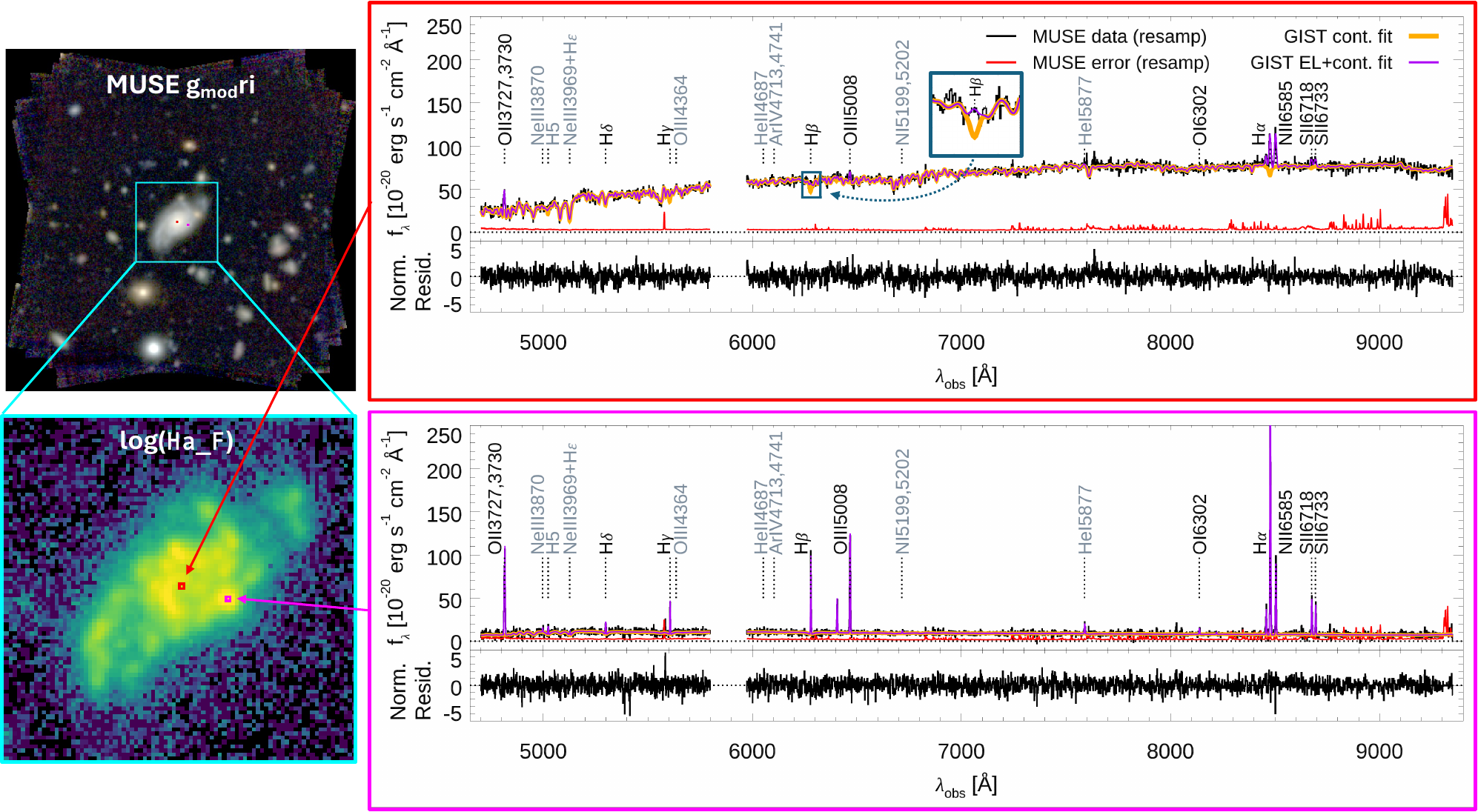}
 \vspace{-1mm}
    \caption{The upper-left panel shows an RGB image of the MAGPI1202 field created from $g_\mathrm{mod}$ (blue), $r$ (green), and $i$ (red) from the MUSE datacube. The lower-left panel is a zoom-in of MAGPI1202197197 showing the Ha flux (\texttt{Ha\_F}). The other panels show the spectral fits and residuals from \gist\ for two spaxels of MAGPI1202197197 in the centre (strong continuum; red) and a star-forming clump (magenta). The MUSE data and error are shown as black and red lines, respectively, and the continuum and continuum+emission line fits are orange and purple lines, respectively. Lines indicated in black text are included in the emission line map products. The upper-right panel includes a zoom-in of the \hbeta\ region to demonstrate that stellar absorption is accounted for in the spectral modelling (i.e., emission line fluxes account for absorption). In all cases, the normalised residuals ($(f_{\lambda,\mathrm{obs}} - f_{\lambda,\mathrm{fit}})/\sigma_{\lambda,\mathrm{obs}}$) are low indicating good fits (resulting in \texttt{REDUCED\_CHI2}<3). The full spectral fits of the datacubes will be made public in a future team data release. 
    \label{fig:GIST_example}}
\end{figure*}

The MAGPI minicubes and aperture-based spectra are fit using the Galaxy IFU Spectroscopy Tool (\gist; \citealt{bittner19}) software\footnote{\url{https://abittner.gitlab.io/thegistpipeline/}}. 
\texttt{GIST} is an all-in-one framework for the analysis of fully reduced IFS data. It is a Python wrapper for \texttt{pPXF} \citep{cappellari&emsellem04}, which performs stellar continuum fitting, and \texttt{GandALF} \citep{sarzi06, falcon-barroso06}, which performs emission line fitting. We note that newer versions of \texttt{pPXF} can also perform emission line fitting \citep{cappellari17}.
The workflow for \gist\ is outlined in Figure~1 of \cite{bittner19} and is summarised below, although we note several custom features that have been implemented for MAGPI\footnote{These new modifications are available through the release of \texttt{nGIST} \citep{fraserMcKelvie25_ngist} at \url{https://github.com/geckos-survey/ngist}}. These custom features along with the values adopted for some of the key \gist\ parameters (see Table~1 of \citealt{bittner19}) are described in Appendix~\ref{gist_config}. Example \gist\ spectral fits for one of the MAGPI group centrals are shown in Figure~\ref{fig:GIST_example}.

First, \gist\ prepares the data for analysis in a series of four preparatory modules: 
(1) reading in the reduced datacube (minicubes or aperture spectra in our case),
(2) logarithmic rebinning of the spectra in the wavelength dimension into linear binning in velocity space,
(3) Voronoi binning of the spectra in the spatial dimensions to obtain integrated spectra of at least a minimum threshold signal-to-noise ratio (via routine of
\citealt{cappellari&copin03}),
and (4) preparation of the adopted spectral template library to match the observational data. 
We used three of the four main analysis modules: (1) stellar continuum modelling and (2) star formation histories, both based on \texttt{pPXF}, and (3) emission line modelling, based on \texttt{GandALF}. We do not use the line indices module. 

For the emission line fitting, we assume that all emission lines are kinematically identical and follow a single Gaussian profile but have freely varying amplitudes (although doublet ratios are fixed for certain lines). 
This assumption is reasonable given the moderate spectral (kinematic) resolution of the MUSE data, noting that typical HII regions have line width offsets of $\sim$2~km/s between Balmer and forbidden lines \citep[e.g.,][]{bresolin20}. 
Emission lines are fit with respect to rescaled versions of the stellar continuum fit from the Voronoi bin where each spaxel resides (see Appendix~\ref{gist_config}). In this respect, line fluxes account for stellar continuum absorption, which can be significant for hydrogen series lines (e.g., see \hbeta\ zoom-in in Figure~\ref{fig:GIST_example}).
The 25 emission lines that are fit are indicated in Table~\ref{tab:emission_lines}. 

\begin{figure*}
	\includegraphics[width=0.99\textwidth]{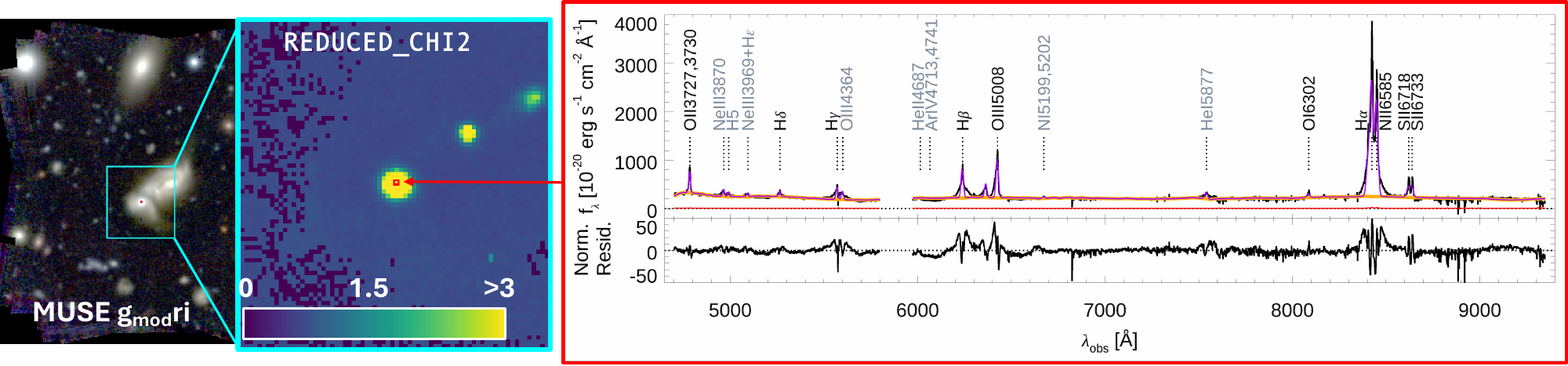}
 \vspace{-1mm}
    \caption{Similar to Figure~\ref{fig:GIST_example} but showing a galaxy with a central AGN region. The middle panel shows the $\chi_\mathrm{red}^2$ map (\texttt{REDUCED\_CHI2}; see eq~\ref{eq:red_chi2}), where values in the AGN region (also centres of neighbouring sources) are poorly fit with a single Gaussian component (\texttt{REDUCED\_CHI2}>3) and line measurements in such regions should be excluded in user analyses. The right panel shows the spectral fits and residuals from \gist\ for the central spaxel of MAGPI2310167176, where a broad component is apparent in the emission lines.
    \label{fig:GIST_AGN_example}}
\end{figure*}

The adoption of single-component Gaussians and the kinematic restrictions are intended to improve the reliability of fits pushing to our faintest limits, but we note that they can be problematic for rare, extreme systems with significant kinematic misalignment between the stellar and gas components (although no cases exceeding the allowed ranges are evident) and/or very broad or multi-component emission line regions 
(e.g., active galactic nuclei, AGN). An example of a source with a poor single-component fit in the central region due to very broad emission (presumably from an AGN) is shown in Figure~\ref{fig:GIST_AGN_example}.

We also run \gist\ on aperture-based (1D) spectra of the 2,607 galaxies with secure redshifts. The type of apertures adopted will be described in Section~\ref{aperture_products}. 

\begin{table*}
	\centering
	\caption{Emission lines included in the \gist\ runs. Wavelengths are in vacuum. Line doublets where a fixed ratio is adopted are indicated. The approximate redshift window(s) for each line are listed, where some have breaks due to the notch filter for the GLAO laser. We indicate the number of galaxies, $N$, for which each line is detected at $S/N\geq3$ and $S/N\geq5$ in the \textit{integrated} spectrum within the undilated \profound\ mask. Lines in bold are included as resolved emission line map products in this data release, with the [OII] doublet provided as a sum. Values of the average Milky Way extinction curve, $k(\lambda_\mathrm{rest})$, from \citet{fitzpatrick19} are provided for making dust corrections of emission lines using \EBVgas\ (see Section~\ref{map_products}).}
	\label{tab:emission_lines}
	\begin{tabular}{lcccccc}
		\hline
		Line & $\lambda_\mathrm{rest}$ (\AA) & Doublet ratio & Redshift window(s)$^*$ & $N(S/N\geq3)$ & $N(S/N\geq5)$ & $k(\lambda_\mathrm{rest})$ [F19]\\
		\hline
        \textbf{[OII]}   & 3727.09 & -- & $0.262-0.556$; $0.602-1.508$ &        1965  &        1783 & 4.707\\
        \textbf{[OII]}   & 3729.79 & -- & $0.261-0.555$; $0.601-1.507$ &        2032  &        1901 & 4.704\\ 
        {[NeIII]} & 3869.79 & -- & $0.215-0.498$; $0.543-1.416$ &        1085  &         639 & 4.553\\
        H5      & 3890.15 & -- & $0.209-0.490$; $0.535-1.403$ &        1062  &         514 & 4.534\\
        {[NeIII]} & 3968.52 & -- & $0.185-0.461$; $0.505-1.356$ &         281  &          72 & 4.469\\
        H$\epsilon$      & 3971.19 & -- & $0.184-0.460$; $0.504-1.354$ &         746  &         331 &  4.467\\
        \textbf{H$\boldsymbol{\delta}$}  & 4102.89 & -- & $0.146-0.413$; $0.456-1.279$ &        1118  &         693 & 4.347\\
        \textbf{H$\boldsymbol{\gamma}$}      & 4341.68 & -- & $0.083-0.335$; $0.376-1.153$ &        1442  &        1098 & 4.148\\
        {[OIII]}  & 4364.38 & -- & $0.077-0.328$; $0.368-1.142$ &         138  &          30 & 4.130\\
        HeII    & 4687.05 & -- & $0.003-0.237$; $0.274-0.995$ &         40  &           1 & 3.818\\
        {[ArIV]}  & 4712.62 & -- & $0.000-0.230$; $0.267-0.984$ &         32  &           4 & 3.800\\
        {[ArIV]}  & 4741.43 & -- & $0.000-0.223$; $0.260-0.972$ &         84  &          11 & 3.779\\
        \textbf{H$\boldsymbol{\beta}$}      & 4862.68 & -- & $0.000-0.192$; $0.228-0.923$ &        1437  &        1274 & 3.687\\
        {[OIII]}  & 4960.21 & 0.350 & $0.000-0.169$; $0.204-0.885$ & -- & -- & -- \\
        \textbf{[OIII]}  & 5008.17 & 1.000 & $0.000-0.158$; $0.193-0.867$ &        1523  &        1405 & 3.530\\
        {[NI]}    & 5199.35 & -- & $0.000-0.115$; $0.149-0.798$ &         71  &          13 & 3.349\\
        {[NI]}    & 5201.84 & -- & $0.000-0.114$; $0.148-0.797$ &         119  &          15 & 3.347\\
        HeI     & 5877.23 & -- & $0.016-0.591$ &         269  &         129 & 2.882\\
        \textbf{[OI]}    & 6301.94 & 1.000 & $0.000-0.483$ &         274  &         155 & 2.666\\
        {[OI]}    & 6365.43 & 0.333 & $0.000-0.469$ & -- & --  & -- \\
        {[NII]}   & 6549.77 & 0.340 & $0.000-0.427$ & -- & --  & -- \\
        \textbf{H$\boldsymbol{\alpha}$}      & 6564.61 & -- & $0.000-0.424$ &         596  &         558 & 2.527\\
        \textbf{[NII]}   & 6585.16 & 1.000 & $0.000-0.420$ &         439  &         317 & 2.516\\
        \textbf{[SII]}   & 6718.16 & -- & $0.000-0.391$ &         409  &         305 & 2.443\\
        \textbf{[SII]}   & 6732.54 & -- & $0.000-0.388$ &         347  &         247 & 2.435\\
		\hline
	\end{tabular}
\newline \textbf{Notes:}  
$^*$The exact redshift window(s) for each line depends on the galaxy kinematics and can be determined by combining the assigned spectroscopic redshift (used for systemic velocity), line velocity, and sigma maps. 
\end{table*}

\section{MAGPI Emission Line Products}\label{EL_products} 
All data products described in this section are made available through the Data Central archive\footnote{\url{https://docs.datacentral.org.au/magpi/}} and in the Supplementary Data of this paper. We note that any version updates to data products will only occur on the Data Central archive and those provided in the Supplementary data are static (and is the version used in the analysis of this paper). Please refer to the Data Central documentation to determine whether any version updates have occurred.

\subsection{Resolved Emission Line Map Products}\label{map_products}

\begin{figure*}
	\includegraphics[width=\textwidth]{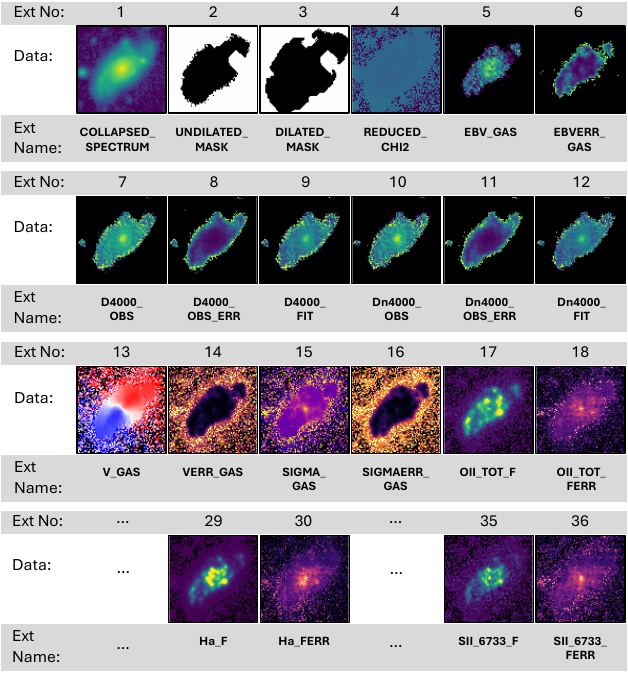}
 \vspace{-4mm}
    \caption{Visual representation of the \texttt{FITS} extensions for the galaxy MAGPI1202197197. Only a subset of the emission line maps are shown for brevity. All maps are shown in linear-scaling except for the first extension, which is shown in log-scaling. For more details on each extension, refer to Table~\ref{tab:FITS_line_maps} and Section~\ref{map_products}.
    \label{fig:Pictoral_EmissionLine_Products}}
\end{figure*}

Emission line maps for 836 individual galaxies at $0.05\leq z_\mathrm{spec}\leq0.424$ (MUSE \halpha\ cutoff) are presented as multi-extension \texttt{FITS} files. Galaxies at $z_\mathrm{spec}>0.424$ are typically unresolved at the spatial resolution of MUSE ($\mathrm{PSF}\sim0.55$\arcsec), and therefore only integrated products are provided for these sources (Section~\ref{aperture_products}). 
The line map products are provided without regard to any emission line detection criterion, and, therefore, there are cases present with undetected line maps (this is common for early-type galaxies).
The $x$- and $y$-dimension sizes of each \texttt{FITS} extension follow exactly the `minicubes' that are input to \gist. 
Each \texttt{FITS} file has 36 extensions that are described in an associated \texttt{README} file in this data release and are summarised in Table~\ref{tab:FITS_line_maps} and a visual representation is provided in Figure~\ref{fig:Pictoral_EmissionLine_Products}. We provide important details on certain extensions below.

\begin{table*}
\caption{Description of MAGPI Emission Line Map Data Products (\texttt{FITS} file extensions). See \texttt{README\_MAGPI\_ELDR\_Map\_products.txt} for more details. \label{tab:FITS_line_maps}}
\begin{tabular}{lll}
\hline \hline
Extension ID & Extension Name & Description \\ \hline

0 & - & Primary header, contains general galaxy information (null data array) \\
1 & \texttt{COLLAPSED\_SPECTRUM} & MUSE white light image (sum) [$10^{-20}$ erg s$^{-1}$ cm$^{-2}$ \AA$^{-1}$] \\
2 & \texttt{UNDILATED\_MASK} & Undilated segmentation mask from \profound, 0=inside mask 1=outside mask \\
3 & \texttt{DILATED\_MASK} & Dilated segmentation mask from \profound, 0=inside mask 1=outside mask \\
4 & \texttt{REDUCED\_CHI2} & $\chi_\mathrm{red}^2$ of the \texttt{GIST} fit (see Eq.~\ref{eq:red_chi2}) \\
5 & \texttt{EBV\_GAS} & $E(B-V)_\mathrm{gas}$ based on the Balmer decrement (see Eq.~\ref{eq:EBVgas}); use this for line dust corrections [mag] \\
6 & \texttt{EBVERR\_GAS} & $E(B-V)_\mathrm{gas}$ $1\sigma$ uncertainty [mag] \\
7 & \texttt{D4000\_obs} & Wide 4000\AA\ break value based on the observed spectrum (see Eq.~\ref{eq:D4000}) \\
8 & \texttt{D4000\_obs\_ERR} & Wide 4000\AA\ break $1\sigma$ uncertainty based on the observed spectrum \\
9 & \texttt{D4000\_fit} & Wide 4000\AA\ break value based on the \texttt{GIST}/\texttt{GandALF} continuum fit \\
10 & \texttt{Dn4000\_obs} & Narrow 4000\AA\ break value based on the observed spectrum (see Eq.~\ref{eq:D4000}) \\
11 & \texttt{Dn4000\_obs\_ERR} & Narrow 4000\AA\ break $1\sigma$ uncertainty based on the observed spectrum \\
12 & \texttt{Dn4000\_fit} & Narrow 4000\AA\ break value based on the \texttt{GIST}/\texttt{GandALF} continuum fit \\
13 & \texttt{V\_GAS} & Ionised gas velocity from \texttt{GIST}/\texttt{GandALF} fit (kinematics tied, uses single Gaussian) [km/s] \\
14 & \texttt{VERR\_GAS} & Ionised gas velocity $1\sigma$ uncertainty (based on MC approach; see Section~\ref{EL_pipeline}) [km/s] \\
15 & \texttt{SIGMA\_GAS} & Ionised gas velocity dispersion from \texttt{GIST}/\texttt{GandALF} fit (kinematics tied, uses single Gaussian) [km/s] \\
16 & \texttt{SIGMAERR\_GAS} & Ionised gas velocity dispersion $1\sigma$ uncertainty (based on MC approach; see Section~\ref{EL_pipeline}) [km/s] \\
17, 19, ..., 33, 35 & \texttt{[LINE]}$^a$ & Emission line flux from \texttt{GIST}/\texttt{GandALF} corrected for MW foreground dust [$10^{-20}$ \esc] \\
18, 20, ..., 34, 36 & \texttt{[LINE]\_err}$^a$ & Emission line flux 1$\sigma$ uncertainty corrected for MW foreground dust [$10^{-20}$ \esc] \\
\hline
\end{tabular} \\
\textbf{Notes:}  
$^a$\texttt{[LINE]} values are, in order: \texttt{OII\_TOT}, \texttt{Hd}, \texttt{Hg}, \texttt{Hb}, \texttt{OIII\_5008}, \texttt{OI\_6302}, \texttt{Ha}, \texttt{NII\_6585}, \texttt{SII\_6718}, \texttt{SII\_6733}. Lines with fixed doublet ratios have only one entry and the flux of the second line can be determined using the ratio listed in Table~\ref{tab:emission_lines}.
We recommend users restrict any analyses to spaxels inside either mask (\texttt{UNDILATED\_MASK}=0 or \texttt{DILATED\_MASK}=0) and \texttt{REDUCED\_CHI2}<3. We also recommend users adopt `Extension Name' strings for reading in parameters in the event that extensions are added and/or the order is modified in future updates. 
\end{table*}

The segmentation masks from \profound\ (\texttt{UNDILATED\_MASK} and \texttt{DILATED\_MASK}; values of 0 and 1 correspond to inside and outside the masks, respectively) are useful for restricting data to individual galaxies of interest, particularly in crowded regions or instances of overlapping galaxies. However, using these masks can exclude regions of the galaxy that may be of interest for emission line measurements. An example of this is apparent in MAGPI1202197197, shown in Figure~\ref{fig:Pictoral_EmissionLine_Products}. There are two additional sources overlapping with this primary galaxy in the upper-right of the 2D maps (this is more evident looking at the \texttt{DILATED\_MASK} extension). These other two sources lie at higher redshifts (both $z_\mathrm{spec}=0.55$) than the primary ($z_\mathrm{spec}=0.29$) such that the emission line fluxes and kinematics of the primary can still be well-recovered (see \texttt{V\_GAS} and \texttt{Ha\_F} panels), although the continuum fit is less accurate in these regions because the stellar continuum is a blend of two sources (the fitting quality can be assessed using the \texttt{REDUCED\_CHI2} extension discussed below). By default, we recommend using only spaxels that reside within either mask (\texttt{UNDILATED\_MASK}=0 or \texttt{DILATED\_MASK}=0) in any analysis. Users who wish to use line measurements in regions where overlapping sources occur (\texttt{UNDILATED\_MASK}=1 or \texttt{DILATED\_MASK}=1) should only do so on a case-by-case basis after considering the redshifts of neighbouring sources (e.g., using Figureset~\ref{fig:MAGPI_environment_examples}) and the \texttt{REDUCED\_CHI2} maps.

The quality of the \gist\ fits are determined based on the $\chi^2$ metric \citep{bevington&robinson03},
\begin{equation}
    \chi^2=\sum_\lambda\frac{(f_{\lambda,\mathrm{obs}} - f_{\lambda,\mathrm{fit}})^2}{\sigma_{\lambda,\mathrm{obs}}^2}  \,,
	\label{eq:chi2}
\end{equation}
and normalised with respect to the number of spectral (i.e., wavelength) channels, $n_\mathrm{fit}$, to get a `reduced'-$\chi^2$, 
\begin{equation}
    \chi_\mathrm{red}^2 =\frac{\chi^2}{n_\mathrm{fit}-n_\mathrm{param}} \simeq \frac{\chi^2}{n_\mathrm{fit}}\,,
	\label{eq:red_chi2}
\end{equation}
where $n_\mathrm{fit}$ is different from the number of MUSE spectral channels due to \gist\ log-rebinning the data into linear velocity space, and $n_\mathrm{param}$ refers to the number of model fitting parameters. In our case, $n_\mathrm{fit}>>n_\mathrm{param}$ such that the values of $\chi_\mathrm{red}^2$ should not differ significantly when normalising only by $n_\mathrm{fit}$ (adopted since $n_\mathrm{param}$ in \gist\ is not well-defined). 
The $\chi_\mathrm{red}^2$ map is provided as the extension \texttt{REDUCED\_CHI2}. The median of all spaxels inside dilated masks and with $S/N(\mathrm{H}\alpha)\geq3$ is \texttt{REDUCED\_CHI2}=0.70 with a standard deviation of 0.47. 
We recommend only using spaxels for which \texttt{REDUCED\_CHI2}<3 (corresponding to $5\sigma$ of the distribution). Values higher than this are instances where the single-component fitting approach we adopt is not suitable (see Figure~\ref{fig:GIST_AGN_example}).

We correct for \textit{foreground} Milky Way (MW) dust extinction in the emission line flux and error maps. We use the average colour excess of the MW foreground, $E(B-V)_{\mathrm{MW}}$, of each MAGPI field from the IPAC dust reddening maps\footnote{\url{https://irsa.ipac.caltech.edu/applications/DUST/}} based on \cite{schlafly&finkbeiner11}, and these values are provided in the \texttt{EBV\_MW} keyword in the primary header of each \texttt{FITS} file for reference (i.e., the same $E(B-V)_{\mathrm{MW}}$ value applies to all galaxies residing in a given MAGPI field). Due to the MAGPI fields residing far from the MW disk, these reddening values are small, ranging from $0.011\leq E(B-V)_{\mathrm{MW}}\leq 0.055$~mag, and usually have only a minor influence on line fluxes (e.g., \halpha\  changes by $\lesssim$10\% for galaxies at $z\sim0.3$). The MW dust-corrected emission line flux is
\begin{equation}
\begin{aligned}\label{eq:MW_line_corr}
F_\mathrm{MWcorr}=F_\mathrm{obs} 10^{(0.4\, k(\lambda_\mathrm{rest}\times(1+z_\mathrm{spec}))\,E(B-V)_{\mathrm{MW}})} \,,
\end{aligned}
\end{equation}
where we use the average MW dust extinction curve, $k(\lambda)$ from \cite{fitzpatrick19} and the \textit{observed} wavelength of each line. 
These correction factors are provided as the \texttt{MW\_CORR} keyword value in the header of each line flux and error map (values differ for each line). If desired, users can divide the values in each map by these factors to remove these MW corrections (i.e., $F_\mathrm{obs}=F_\mathrm{MWcorr}/$\texttt{MW\_CORR}).

We also provide a map of the colour excess of the ionised gas, \EBVgas, (\texttt{EBV\_GAS}) for users to perform \textit{internal} dust corrections on emission lines.
The \EBVgas, is derived from the Balmer decrement \textit{after correcting} for MW foreground dust reddening,
\begin{equation}
\begin{aligned}\label{eq:EBVgas}
E(B-V)_{\mathrm{gas}} = \frac{\log((F(\mathrm{H}\alpha)/F(\mathrm{H}\beta))/2.863)}{0.4(k(\mathrm{H}\beta)-k(\mathrm{H}\alpha))}  \,,
\end{aligned}
\end{equation}
where 2.863 is the theoretical value expected for the unreddened ratio of $F(\mathrm{H}\alpha)/F(\mathrm{H}\beta)$ undergoing Case B recombination with $T_{\mathrm{e}}=10^4$~K and $n_{\mathrm{e}}=100$~cm$^{-3}$ \citep{osterbrock89,osterbrock&ferland06}\footnote{In Case B recombination, the theoretical ratios for $F(\mathrm{H}\alpha)/F(\mathrm{H}\beta)$ change by $\lesssim$10\% over the range of physical conditions expected for typical HII regions \citep{osterbrock&ferland06} and therefore the exact properties assumed have a small influence. The assumption of Case B is expected to be valid for typical star forming galaxies but could be incorrect for extreme emission line galaxies \citep[e.g.,][]{scarlata24}.}, and we assume an average MW extinction curve, $k(\lambda)$, at the wavelengths of \hbeta\ and \halpha, with $k(\mathrm{H}\beta)-k(\mathrm{H}\alpha)=1.160$ \citep{fitzpatrick19}. Estimates of \EBVgas\ are only provided for spaxels where $S/N(\mathrm{H}\alpha) \ge 3$ and $S/N(\mathrm{H}\beta) \ge 3$. The cases where negative values of \EBVgas\ are inferred are set to \texttt{EBV\_GAS=0}. 

The \EBVgas\ can be used to make dust attenuation corrections of the MW dust corrected emission line flux maps,
\begin{equation}
\begin{aligned}\label{eq:line_corr}
F_\mathrm{corr}=F_\mathrm{MWcorr} 10^{(0.4\,k(\lambda_\mathrm{rest})\,E(B-V)_{\mathrm{gas}})} \,.
\end{aligned}
\end{equation}
The MW extinction curve has been found to be representative of the average nebular dust attenuation curve for massive star-forming galaxies in the local Universe \citep[$z\sim0$; e.g.,][]{rezaee21} and at $z\sim2$ \citep[e.g.,][]{reddy20}. For this reason, we recommend users to adopt the values of the average MW extinction curve $k(\lambda)$ provided in Table~\ref{tab:emission_lines} \citep{fitzpatrick19} as the default choice for dust corrections of emission lines.

We include measurements of two metrics for the 4000\AA\ break feature, the `wide' version  \citep[$D(4000)$;][]{bruzual83} and the `narrow' version \citep[$D_n(4000)$;][]{balogh99, kauffmann03a}, which are defined as:
\begin{equation}
    D(4000)=\frac{\langle f_{\nu}(4050\textup{\AA}-4250\textup{\AA}) \rangle }{\langle f_{\nu}(3750\textup{\AA}-3950\textup{\AA}) \rangle }\,,
	\label{eq:D4000}
\end{equation}
\begin{equation}
    D_n(4000)=\frac{\langle f_{\nu}(4000\textup{\AA}-4100\textup{\AA}) \rangle }{\langle f_{\nu}(3850\textup{\AA}-3950\textup{\AA}) \rangle }\,.
	\label{eq:Dn4000}
\end{equation}
The narrow version has the benefit that it is less affected by dust attenuation and also experiences a smaller redshift gap due to the notch filter for the VLT GLAO laser. However, the wider version can be beneficial when the continuum signal-to-noise ratio is low. 
We require the entire window to be covered by the MUSE spectrum and that each window have $S/N\ge2$ to estimate each parameter. The redshift gap due to the notch filter is $0.365\leq z \leq 0.592$ and $0.415 \leq z \leq 0.551$ for $D(4000)$ and $D_n(4000)$, respectively. The values are provided using the observed MUSE data directly (e.g., \texttt{D4000\_OBS}) or from the stellar continuum fit (e.g., \texttt{D4000\_FIT}). The latter is less prone to noise issues (e.g., poor skyline subtraction) but is also dependent on the stellar models being reliable. 

Due to the fact that the [OII] doublet is blended together at the MUSE spectral resolution for most of our sample (unresolved for $z\lesssim0.9$), we only provide a summed flux of both lines \texttt{OII\_TOT}. This is adopted because with the freely-varying flux amplitudes of the two lines, there is significant degeneracy on the flux assigned to each, and comparisons of their line ratios should not be trusted.

We provide two ascii tables that summarise the fraction of spaxels within the \profound\  \textit{undilated} mask that are detected with $S/N\geq 3$ and $S/N\geq 5$ in each emission line listed in Table~\ref{tab:emission_lines}. We adopt the undilated mask because the outskirts included in the dilated mask are often not detected in the emission lines. These files are named \texttt{MAGPI\_ELDR\_SNRge3\_emission\_line\_frac.csv} and \texttt{MAGPI\_ELDR\_SNRge5\_emission\_line\_frac.csv}. These tables include a column labelled \texttt{flag} that indicates the continuum $S/N$ threshold adopted for the Voronoi binning of each source (see Section~\ref{EL_pipeline}). There is also a column labelled \texttt{N\_MASK} that indicates the number of spaxels residing in the dilated mask. Multiplying \texttt{N\_MASK} by the detected fraction for each line gives the number of spaxels detected for that line.

\begin{figure*}
	\includegraphics[width=0.8\textwidth]{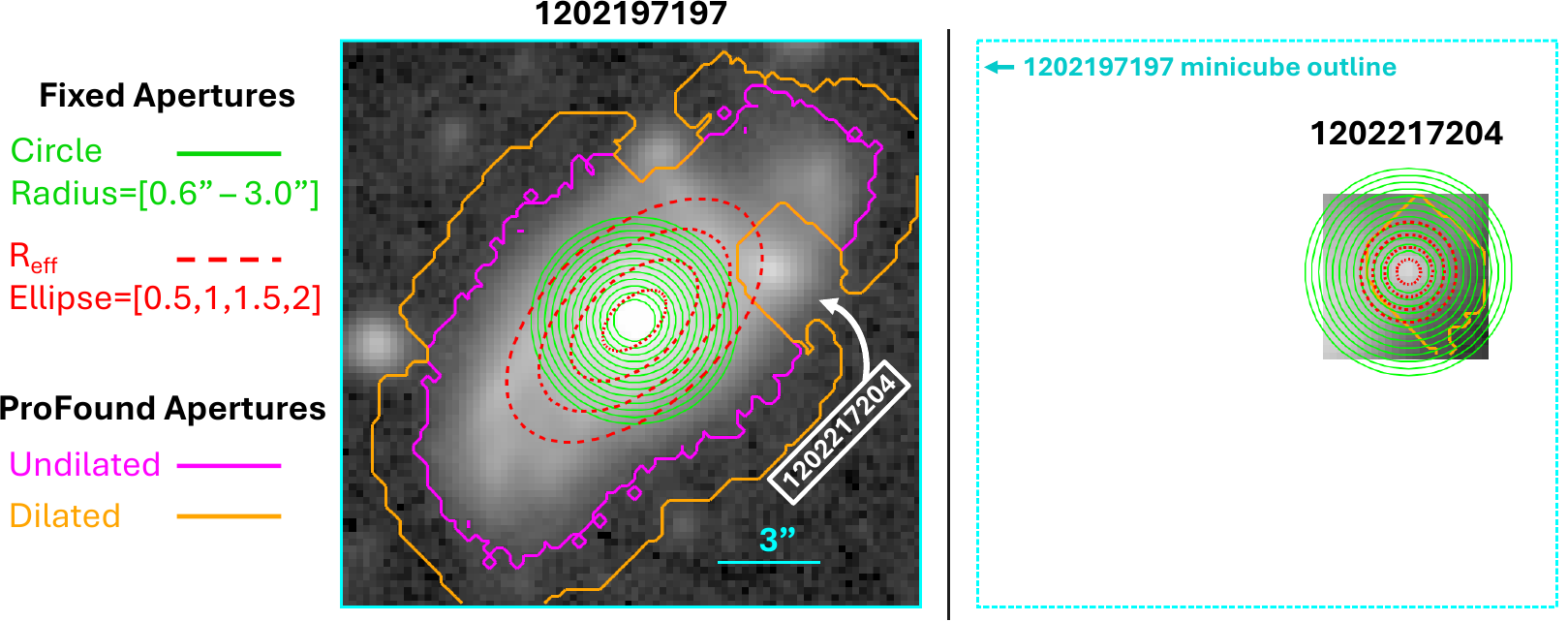}
 \vspace{-3mm}
    \caption{Visual representation of the \texttt{ondespec} apertures for the galaxies MAGPI1202197197 (left) and MAGPI1202217204 (right), which neighbour each other. The apertures are overlayed on top of the \texttt{COLLAPSED\_SPECTRUM} extension from the minicubes. MAGPI1202197197 is among the largest angular-sized star-forming galaxies in the survey. 
    There are thirteen fixed circular apertures with radii from 0.6\arcsec\ to 3.0\arcsec\ in increments of 0.2\arcsec, four $R_e$-based apertures [0.5$R_e$, 1.0$R_e$, 1.5$R_e$, 2.0$R_e$], and two \profound\ mask-based apertures [dilated, undilated].
    The fluxes from the fixed apertures are measured from the sum of \textit{only spaxels residing inside the dilated mask} (orange contour). For MAGPI1202197197, only a portion of spaxels inside the 2$R_\mathrm{eff}$ aperture that overlaps with part of MAGPI1202217204 would get excluded. However, a much larger number of spaxels are excluded in the fixed-apertures for MAGPI1202217204. As a result, the line fluxes from the largest fixed-apertures in MAGPI1202217204 will all have similar values. For MAGPI1202217204, the undilated and dilated aperture extents are identical. We recommend using the undilated or dilated apertures for integrated line flux measurements.
    \label{fig:Aperture_examples}}
\end{figure*}

\subsection{Aperture Emission Line Products}\label{aperture_products}
We provide a catalogue of emission line measurements using 1D spectra (hereafter `\texttt{onedspec} products') for a variety of apertures as a \texttt{FITS} binary table, which is named \texttt{MAGPI\_ELDR\_onedspec\_products.fits}. 
This table includes all 2,607 galaxies with secure redshifts at $z\leq1.5$ and \textit{all} emission lines fit with \gist\ (see Table~\ref{tab:emission_lines}).
Measurements are provided for three types of apertures: (1) fixed angular size, (2) effective size, and (3) within \profound\ masks. 
There are thirteen fixed circular apertures with radii from 0.6\arcsec\ to 3.0\arcsec\ in increments of 0.2\arcsec, four $R_e$-based elliptical apertures [0.5$R_e$, 1.0$R_e$, 1.5$R_e$, 2.0$R_e$], and two \profound\ mask-based apertures [dilated, undilated]. 
The axial ratio of the elliptical apertures is fixed to the global value derived from \profound. 
A visualisation of the apertures for two neighbouring galaxies is shown in Figure~\ref{fig:Aperture_examples}. 
This table and a corresponding \texttt{README} of the entries are provided on Data Central and in the Supplementary Data of this paper. 

We correct all emission line fluxes and errors for foreground MW dust extinction using the same method described in Section~\ref{map_products}. We provide the same derived properties for the resolved products in the \texttt{onedspec} products (\EBVgas\ and $D4000$). As before, estimates of \EBVgas\ are provided only for apertures where $S/N(\mathrm{H}\alpha) \ge 3$ and $S/N(\mathrm{H}\beta) \ge 3$ and the cases where negative values are inferred are set to \texttt{EBV\_GAS=0}. 

The \profound\ mask-based apertures are likely to be the most useful for performing sample selections because they provide integrated galaxy measurements (with the caveat that these can underestimate the extent of overlapping sources). We note that the dilated mask is a larger area than the undilated mask and hence may be more representative of the `total' galaxy area, however, the $S/N$ of emission lines in galaxy outskirts tend to be low such that the integrated $S/N$ of lines measured in the dilated mask tend to be lower than the undilated mask. A summary of the fraction of sources with detections of each emission line measured with \gist\ in the undilated mask is listed in Table~\ref{tab:emission_lines}. 
The 0.6\arcsec\ radius circular aperture is recommended to characterise whether a galaxy has a central dominant AGN (e.g., via line ratio diagnostics).

\subsection{Integrated Star Formation Rate Catalogue}\label{SFR_products}
We provide two ascii tables of total SFRs inferred from dust-corrected \halpha\ via Balmer line ratios (\texttt{SFR\_Hacorr}), measured within the \profound\ undilated and dilated mask (i.e., \texttt{onedspec} product described in Section~\ref{aperture_products}), which are named \texttt{MAGPI\_ELDR\_Balmer\_SFR\_undilated.csv} and \texttt{MAGPI\_ELDR\_Balmer\_SFR\_dilated.csv}. These SFR estimates are simplistic and do not attempt to correct or remove sources with AGN contamination in the emission lines. 
The Balmer line ratio used for estimating \EBVgas\ depends on the source redshift. In order of decreasing preference, we use \halpha/\hbeta, \halpha/H$\gamma$, \hbeta/H$\gamma$, and \hbeta/H$\delta$ (see Table~\ref{tab:emission_lines} for emission line windows/gaps). Assuming Case B recombination with $T_{\mathrm{e}}=10^4$~K and $n_{\mathrm{e}}=100$~cm$^{-3}$, the theoretical values expected for the unreddened ratios of these lines are 2.863, 6.118, 2.137, and 3.861, respectively \citep{osterbrock89,osterbrock&ferland06}. 
These tables and a corresponding \texttt{README} of the entries are provided on Data Central and in the Supplementary Data of this paper.

Star formation rates (SFRs) are derived from extinction-corrected \halpha\ using Eq.~(\ref{eq:line_corr}) with the $k(\mathrm{H}\alpha)$ value listed in Table~\ref{tab:emission_lines} and the integrated \EBVgas\ values. 
We convert the extinction-corrected \halpha\ flux to a luminosity based on the luminosity distance from \zspec\ and then use the SFR calibration from \cite{calzetti13}, which assumes a \cite{kroupa01} initial mass function (IMF),
\begin{equation}
\begin{aligned}\label{eq:sfr}
\mathrm{SFR}(M_\odot~\mathrm{yr}^{-1}) = 5.5\times 10^{-42} L(\mathrm{H}\alpha)_\mathrm{corr} \,,
\end{aligned}
\end{equation}
where the \halpha\ luminosity is measured in erg~s$^{-1}$. We note that SFRs derived assuming either the \cite{kroupa01} or \cite{chabrier03} IMFs are nearly identical \citep{kennicutt&evans12} and that this table can be considered suitable for both assumptions. We require that the integrated Balmer lines have $S/N\ge3$ for a SFR to be calculated, and therefore SFRs for passive galaxies are not provided (values are listed as \texttt{NaN}). 

The SFR uncertainties are expressed as 16th and 84th percentiles and include a covariance term to account for the fact that the SFR is being derived from a Balmer line and this same line is also used in the dust correction (i.e., Balmer decrement). This increases the error slightly over a formal error that assumes independent variables. In instances where the SFR uncertainties would imply a 16th percentile that is below zero (typically only occurs when using the higher-order Balmer ratios), we use the dust-uncorrected line flux minus the dust-uncorrected 1$\sigma$ flux uncertainty to determine it, which can be considered the lower limit. 
The Balmer line ratio adopted is indicated by the value of \texttt{SFR\_Hacorr\_FLAG} with the use of \halpha/\hbeta, \halpha/H$\gamma$, \hbeta/H$\gamma$, and \hbeta/H$\delta$ indicated by values 0, 1, 2, and 3, respectively.

\section{Sample Selection for Results}\label{sample_selection}
We make a series of additional selection cuts to define parent samples for the SAMI ($z\sim0$ benchmark) and MAGPI data used in the results in Section~\ref{results}. First, we require that all galaxies have integrated $S/N\geq3$ for \hbeta\ and \halpha. For SAMI, this is based on the sum of all spaxels inside the 15\arcsec\ hexabundle. For MAGPI, we use the emission line measurements within the undilated \profound\ mask from the aperture products (Section~\ref{aperture_products}).

\begin{figure}
	\includegraphics[width=0.48\textwidth]{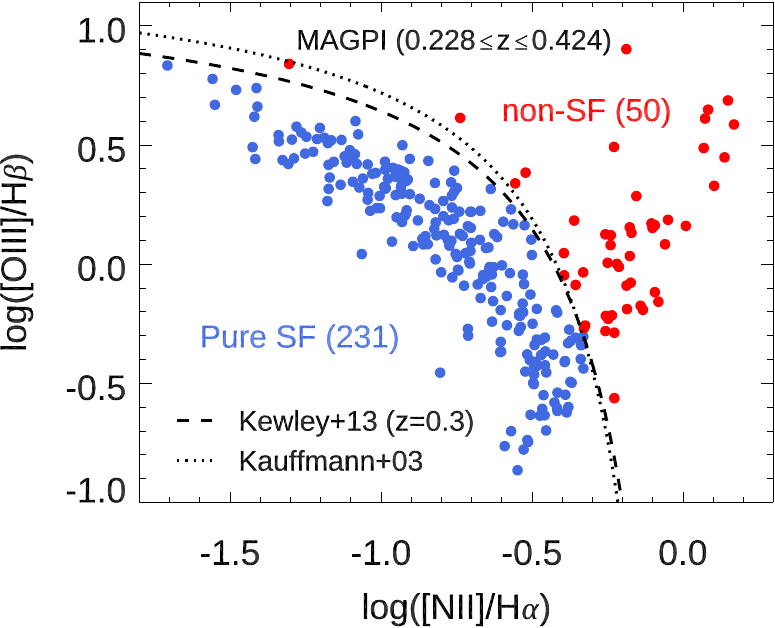}
 \vspace{-4mm}
\caption{BPT diagram for MAGPI galaxies at $0.228\leq z_\mathrm{spec} \leq 0.424$ using flux within the 0.6\arcsec\ radius circular aperture. The dashed line indicates the demarcation line from \citet{kewley13b}  at $z=0.3$ (individual galaxy redshifts are used for actual assignment). The \citet{kewley13b} line at low-$z$ is roughly consistent with the demarcation from \citet[][dotted line]{kauffmann03c}, based on SDSS-III, and would provide a similar selection of star-forming galaxies. Only galaxies in the `Pure SF' region are retained for our results in Section~\ref{BD_radius_compare} and \ref{BD_mstar_compare}.}
\label{fig:MAGPI_BPT}
\end{figure}

We flag galaxies for which the central region is dominated by non-stellar ionisation sources, which can arise from shocks or an active galactic nucleus (AGN). The cases dominated by non-stellar ionisation are determined using the BPT \citep*{baldwin81} diagnostic diagram  and the demarcation of \cite{kewley13b} in [OIII]/\hbeta\ as functions of [NII]/\halpha\ and redshift, which separates ionisation from purely star-forming and non-star-forming sources (shocks and AGN). 
We require sources to have $S/N\geq3$ for \hbeta, [OIII], \halpha, and [NII] in the central region and be located in the `Pure SF' to be used in our Balmer-decrement results (Sections~\ref{BD_radius_compare} and \ref{BD_mstar_compare}). 
For SAMI, the central region is defined as a circular aperture of 3\arcsec\ diameter, with the line measurements being taken directly from the DR3 1-component aperture flux catalogue (\texttt{EmissionLine1compDR3}).
For MAGPI, the central region is defined as a circular aperture of $R=0.6$\arcsec, and the line measurements are taken from the \texttt{onedspec} products. 
We show the distribution of the MAGPI sample at $0.228<z<0.424$ in the BPT diagram in Figure~\ref{fig:MAGPI_BPT}. 

For SAMI, we apply standard quality flag selections to ensure reliable calibration, sky subtraction, and emission-line fitting\footnote{\texttt{WARNSTAR == 0}, \texttt{WARNMULT != 2}, \texttt{ISBEST != False}, \texttt{WARNFCAL == 0}, \texttt{WARNFCBR == 0}, \texttt{WARNSKYB == 0}, \texttt{WARNSKYR == 0}, \texttt{WARNSKEM == 0}, \texttt{WARNEMFT == 0}, \texttt{BADCLASS == 0}.}. 
We exclude SAMI galaxies where the hexabundle covers less than 50\% of the total stellar mass, which is assessed by the ratio of the summed mass in the hexabundle area (from SAMI mass maps) relative to the total stellar mass (from GAMA catalogue). We also limit the sample to $z<0.06$ due to the fact that there are very few galaxies beyond this redshift after meeting the quality criteria above.  
After making these cuts, the SAMI parent sample contains 918 star-forming galaxies (after a BPT cut), spanning $0.004<z<0.059$ and $7.20\leq \log[M_\star/M_\odot]\leq10.91$. The median covering fraction (by mass), $f_c$, of the SAMI hexabundle for this parent sample is $\left<f_c\right>\sim0.86$, indicating that a majority of each galaxy is covered (by design from our selection above). This implies that line ratios determined from the summed flux should be minimally affected by radial gradients.

For MAGPI, we restrict the sample to $0.228\leq z_\mathrm{spec} \leq 0.424$. The lower redshift boundary corresponds to where \hbeta\ leaves the laser gap and the upper boundary is where \halpha\ shifts outside the red-end of the MUSE spectral coverage. Due to the large MUSE FoV, the median covering fraction of MAGPI galaxies can be considered $\left<f_c\right>\sim1$. We also restrict the sample to cases where reliable stellar masses can be determined. After these cuts, the MAGPI parent sample contains 410 galaxies with integrated BD values ($S/N\geq3$ for \hbeta\ and \halpha) and 227 galaxies after BPT cuts ($S/N\geq3$ also for [OIII] and [NII]).

\section{Results}\label{results} 
In the sections below, we provide a basic overview of a subset of the MAGPI galaxies and their application to examine BD radial gradients and the BD-$M_\star$ relation. We highlight that variants of the emission line products in this release have been used to study a wide variety of topics, including the resolved star-forming main sequence \citep{mun24, mun25}, kinematic asymmetries of ionised gas \citep{bagge23, bagge24}, the nature of spiral arms \citep{chen24a}, the resolved SFR-mass-metallicity relation \citep{koller24}, gas turbulence \citep{mai24}, gas-phase metallicity gradients \citep{mai26}, galaxy dark matter fractions \citep{sharma26}, stellar population property gradients \citep{mauro26}, and predicting resolved dust attenuation \citep{mailvaganam26b}. 
We refer readers to these works for examples of applying these data products to study those topics.

\subsection{MAGPI Emission Line Galaxy Sample at $z\sim0.3$ Overview}\label{sample_overview} 
\begin{figure}
	\includegraphics[width=0.48\textwidth]{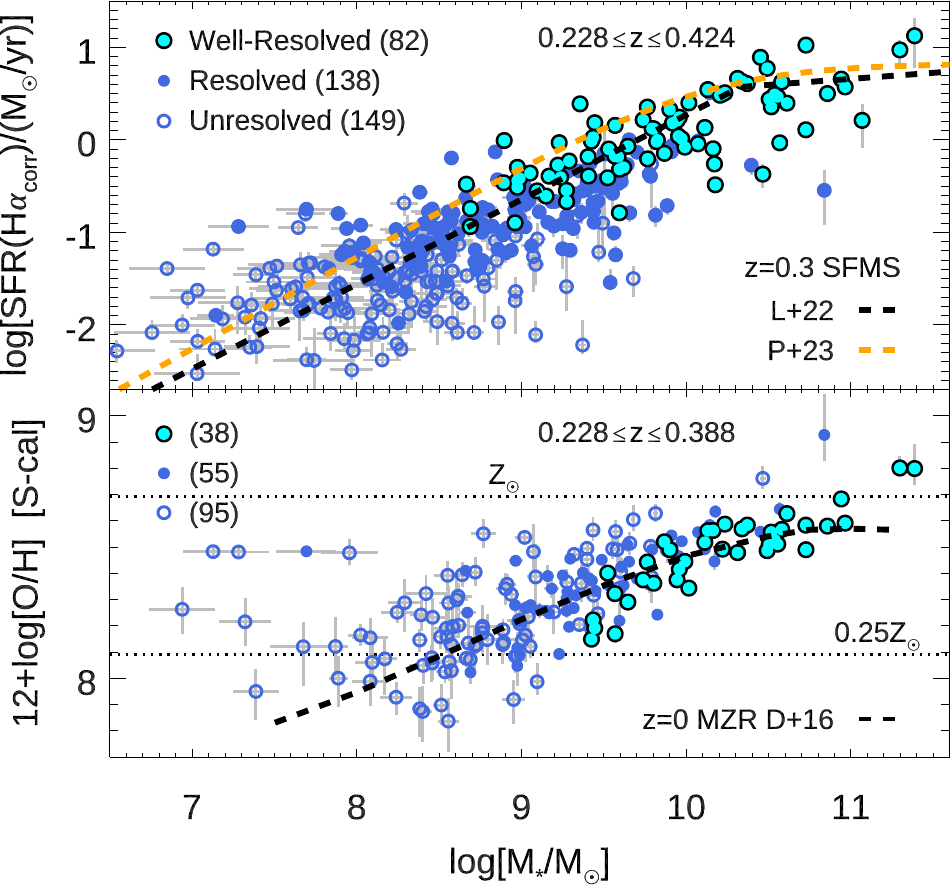}
 \vspace{-4mm}
\caption{\textit{Top:} SFR from Dust-corrected \halpha\ as a function of stellar mass, also known as the star-forming galaxy main sequence (SFMS), for MAGPI galaxies at $0.228\leq z_\mathrm{spec} \leq 0.424$. The dashed coloured lines indicate the SFMS relationships at $z\sim0.3$ from \citet{leja22} and \citet{popesso23}.  \textit{Bottom:} Gas-phase metallicity (based on $S$-cal, \citealt{pilyugin&grebel16}) as a function of stellar mass, also known as the mass-metallicity relation (MZR), for MAGPI galaxies at $0.228\leq z_\mathrm{spec} \leq 0.388$. The upper boundary is lower due to requiring [SII], which shifts out of MUSE earlier than \halpha. The dashed curve indicates the MZR relationship at $z\sim0$ based on $S$-cal from \citet{deVis19} derived from the DustPedia sample. 
In both panels, we distinguish cases that are considered well-resolved, resolved, and unresolved in the emission lines used for deriving the quantities shown on the y-axes (see Section~\ref{sample_overview} for definitions). Cases where the central region is dominated by non-stellar ionisation (see Figure~\ref{fig:MAGPI_BPT}), are excluded in both panels, but the top panel retains sources that are undetected in [OIII] and [NII].}
\label{fig:MAGPI_MS_MZR}
\end{figure}

We present an overview of the parameter space in integrated galaxy stellar mass ($M_\star$), star formation rate (SFR), and gas-phase metallicity of the MAGPI star-forming galaxy sample at $z\sim0.3$ in Figure~\ref{fig:MAGPI_MS_MZR}. We use the SFR measurements within the undilated \profound\ mask from the aperture products (\ref{SFR_products}). 
We explain how the stellar mass and gas-phase metallicities are derived in the following paragraphs.

For the MAGPI fields in the G12 and G15 fields (44/56 fields), $M_\star$ maps have been derived from Subaru HSC $grizy$ bands using the methodology of \cite{taylor11} and will be presented in an upcoming release. 
This approach uses inferred mass-to-light ratios ($M/L$) as a function of $g-r$ colour. We use a bootstrapped relationship between the $g-r$ colour and $M/L$ in the $r$ band from the MUSE data to extend this approach to the entire sample so that we have mass maps consistently across the full 56 fields. For more details, we refer the reader to \cite{mun24} that use the same data. The bootstrapped mass maps are only constructed for sources at $z<0.43$ where the MUSE data covers rest-frame optical continuum and stellar masses based on $M/L$ ratios are more reliable (relative to using rest-frame UV continuum at higher redshifts). We present a comparison of our integrated bootstrap mass estimates and those derived from integrated SED fitting using the \texttt{ProSpect} code \citep{robotham20} for GAMA \citep{bellstedt20a, bellstedt20b}, in Appendix~\ref{mass_compare}, finding that there is good general agreement (median offset of 0.12~dex) although with moderate scatter ($\sigma=0.32$~dex).

We adopt the $S$-calibration ($S$-cal) prescription of \citet{pilyugin&grebel16} to estimate the gas-phase metallicity. This prescription uses a combination of the following line ratios:
\begin{align}\label{eq:another}
\begin{split}
\mathrm{N}_2 &= (\mathrm{[NII]}\lambda 6548 + \lambda 6584) / \mathrm{H}\beta \,,\\
\mathrm{S}_2 &= (\mathrm{[SII]}\lambda 6717 + \lambda 6731) / \mathrm{H}\beta \,,\\
\mathrm{R}_3 &= (\mathrm{[OIII]}\lambda 4959 + \lambda 5007) / \mathrm{H}\beta \,.
\end{split}
\end{align}
These lines are extinction corrected following Eq.~(\ref{eq:line_corr}) and using $k(\lambda)$ values for each line in Table~\ref{tab:emission_lines} and integrated \EBVgas\ values. We require that all emission lines needed in these ratios have integrated $S/N\ge3$ to estimate a global metallicity. The choice of using $S$-cal is driven by it being less biased on the ionisation parameter and also reliable over a very wide metallicity range ($7.0\lesssim12+\log[\mathrm{O/H}]\lesssim8.8$; \citealt{pilyugin&grebel16}) relative to many other calibrations. There can be large systematic offsets between different metallicity diagnostics \citep[e.g.,][]{kewley&ellison08} and this should be taken into account when making comparisons to other samples. Such comparisons are beyond the scope of this work. 
We choose not to provide a catalogue of gas-phase metallicity because of the wide variety in available diagnostics. The emission line products provided in this release can be used to calculate the preferred diagnostic of the user.

As can be seen in Figure~\ref{fig:MAGPI_MS_MZR}, the primary MAGPI sample spans a wide range of properties, covering roughly 5~dex in $M_\star$ and SFR, and 1~dex in gas-phase metallicity. The majority of galaxies are representative of typical star forming galaxies based on comparison to two reference star forming main-sequence relations (SFMS) at $z=0.3$ \citep{leja22, popesso23} and the mass-metallicity (MZR) relation at $z=0.0$ from \cite{deVis19} based on $S$-cal. The latter assertion is based on the evolution of gas-phase metallicity from $z=0.3$ to $z=0.0$ being mild \citep[e.g.,][based on different calibrations than $S$-cal]{maiolino&mannucci19}. 
We take the result that most of the MAGPI star-forming galaxies reside on or near the SFMS as an indication that the environmental selection of MAGPI is not significantly biasing the emission-line-selected sample. For reference, 58\% of the $z\sim0.3$ star-forming sample are within $|\Delta z|\leq0.03$ of group central galaxies. 
The decrease in the sample size with metallicity measurements relative to SFR is due to the requirement to detect more lines (primary factor) and the slightly narrower redshift window (since [SII] is used). Galaxies at high stellar mass are more likely to contain non-stellar ionising sources in their centres (not shown), which is expected since they are more likely to host AGN \citep[e.g.,][]{juneau11}.

In Figure~\ref{fig:MAGPI_MS_MZR}, we distinguish sources with differing amounts of resolved regions available. This distinction is based on the spaxel area with all required emission lines at $S/N\geq3$ for the derived measurement (e.g., \halpha\ and \hbeta\ for extinction-corrected SFRs) relative to the MUSE AO PSF ($\mathrm{FWHM}=0.55$\arcsec). 
This definition differs from using the effective radius in a particular broadband filter, which is based on the stellar continuum and is commonly adopted as the reference size. 
We divide the sample into three categories: `well-resolved', `resolved', and `unresolved' based on the spaxel area of the detected emission lines covering the equivalent of $R\geq 2\mathrm{FWHM}$, $1\mathrm{FWHM}\leq R<2\mathrm{FWHM}$, $R<1\mathrm{FWHM}$, respectively. For $R=1\mathrm{FWHM}$, the area is $\pi\times(0.55\arcsec/0.2\arcsec)^2=23.76$~pixels (for FWHM=0.55\arcsec\ and 0.2\arcsec\ pixels). 
These values correspond to equivalent area coverage of $A\geq16\mathrm{PSF}$, $4\leq A<16\mathrm{PSF}$, $A<4\mathrm{PSF}$, respectively.
The number of spaxels detected for each line is taken from the \texttt{MAGPI\_ELDR\_SNRge3\_emission\_line\_frac.csv} data product.
These demarcations are arbitrary, but help to demonstrate the quality of the data in different areas of parameter space. 
As expected, the most massive (larger average angular size) and highest SFR galaxies are preferentially more resolved in their emission lines. 
There is also a slight preference for more resolved galaxies, which have higher SFRs, to have lower gas-phase metallicities at a fixed stellar mass. This is an expected trend in the fundamental metallicity relation \citep[e.g.,][]{mannucci10} where there is an anti-correlation between SFR and metallicity at fixed stellar mass.

\subsection{Balmer Decrement Radial Gradients from SAMI and MAGPI}\label{BD_radius_compare}
Resolved observations of Balmer decrements from SAMI and MAGPI provide insight into the distribution of dust attenuation throughout galaxies. We use BD maps from these surveys to determine average BD gradients of galaxies in different mass bins, out to $R\sim3R_e$ for SAMI and $R\sim2R_e$ for MAGPI. These allow us to understand the role of aperture choice on Balmer decrement measurements and the BD-$M_\star$ relation. 

Starting from the parent samples for SAMI and MAGPI (see Section~\ref{sample_selection}), we measure BD values by summing flux within elliptical annuli that increase in effective size ($\sqrt{(x-x_c)^2+(y-y_c)^2} $) by 1 spaxel (0.5\arcsec\ for SAMI; 0.2\arcsec\ for MUSE) although we start with an ellipse (not annulus) of radius of 1 spaxel (2 spaxel diameter) for the first step. The elliptical shapes are based on global measurements from \profound\ for each galaxy. We find that the results are qualitatively consistent if we adopt circular instead of elliptical apertures, but that the scatter in BD values at a given radius increases if using circular apertures.  
We restrict this analysis to spaxels that meet the following criteria:
\begin{enumerate}
\item[(1)] $S/N(\mathrm{H}\alpha) \ge 5$ and $S/N(\mathrm{H}\beta) \ge 1$
\item[(2)] Good quality emission line fit
\item[(3)] Uncontaminated by neighbouring galaxies
\end{enumerate}
For condition (1), the low $S/N$ threshold for \hbeta\ ensures that we do not bias ourselves against very dusty regions where \hbeta\ can become much weaker than \halpha. 
For SAMI, condition (2) \& (3) are ensured through the \texttt{WARNEMFT == 0} and \texttt{WARNMULT != 2} quality flags, respectively, used to select the parent sample, as well as by excluding spaxels at $R>4R_e$. 
For MAGPI, conditions (2) \& (3) are ensured through \texttt{REDUCED\_CHI2}<3 and \texttt{DILATED\_MASK}=0, respectively.
Only annuli where at least 10\% of spaxels in the annulus meet the above $S/N$ criteria are retained for both SAMI and MAGPI. 
We further restrict to only retain annuli where the \textit{summed} flux achieves $S/N(\mathrm{H}\alpha)\geq5$ and $S/N(\mathrm{H}\beta)\geq3$, respectively. 
These criteria reduce the SAMI and MAGPI sample to 918 and 177 galaxies, respectively, that contain 20,895 and 1,625 total annuli, respectively, for the results of this Section. 

We divide the sample into bins of stellar mass such that the number of annuli used to measure the gradients in each mass bin is similar (each with $\sim$4180 and $\sim$325 annuli for SAMI and MAGPI, respectively). This means that more unique galaxies contribute to the lower-mass bins than to the higher-mass bins, because more annuli are detected in more massive galaxies. For reference, in SAMI, the lowest and highest mass bins have an average of 18.6 and 26.0  annuli per galaxy, respectively.  In MAGPI, the lowest and highest mass bins have an average of 6.5 and 19.7 annuli per galaxy, respectively.

\begin{figure*}
$\begin{array}{cc}
	\includegraphics[width=0.48\textwidth]{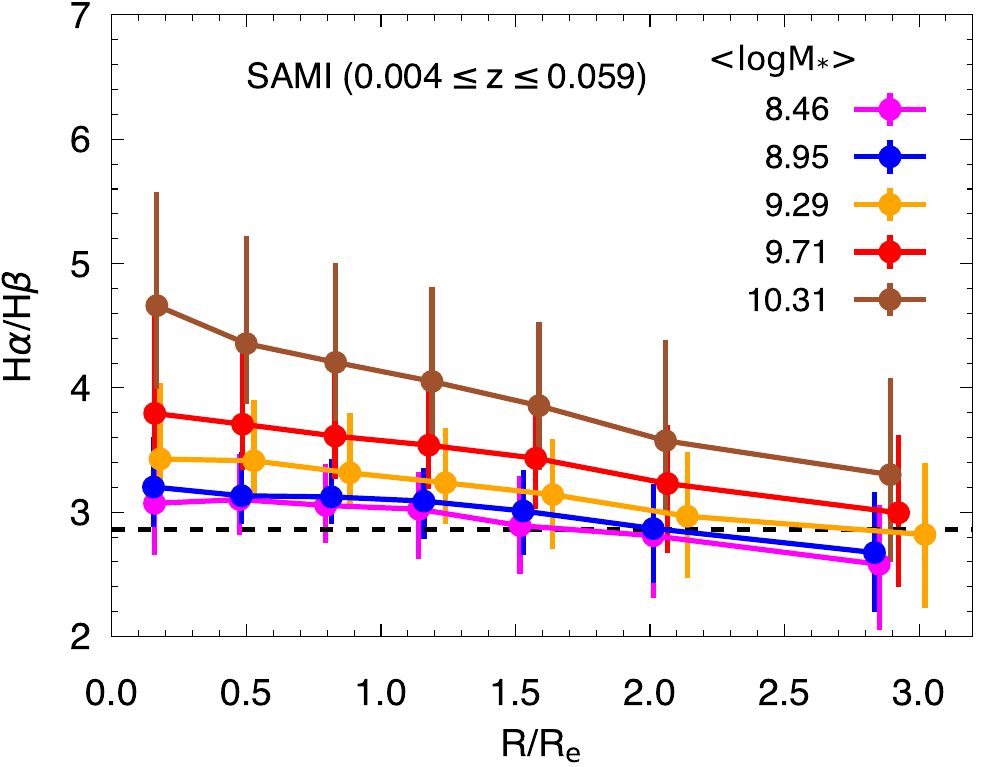} & \hspace{1mm}
\includegraphics[width=0.48\textwidth]{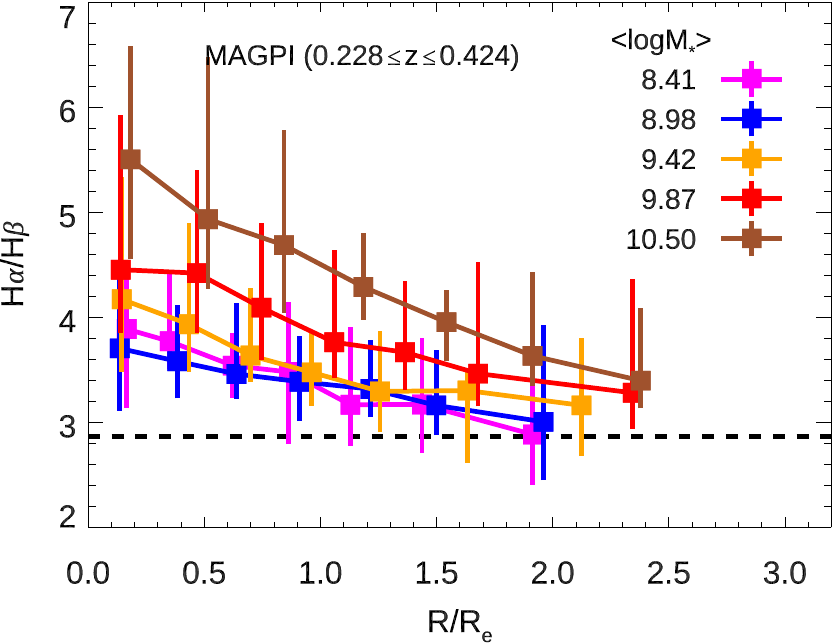} \\
\end{array}$
\vspace{-3mm}
    \caption{Balmer decrement vs radius for galaxies in \textit{(Left:)} SAMI and \textit{(Right:)} MAGPI separated into five mass-bins with roughly equal-number of annuli. These are subdivided into seven equal-number radial bins. The errorbars indicate the 16th and 84th percentiles of the intrinsic scatter and not the measurement uncertainty, where the latter are usually smaller. The trends are similar between samples (redshift), with more massive galaxies having larger BD values and steeper gradients. The MAGPI BDs are systematically higher than the SAMI values of similar mass, suggesting they are on average slightly dustier. The horizontal dashed line is the intrinsic BD ratio (2.863) expected in the absence of dust attenuation.
    \label{fig:SAMI_MAGPI_BD_vs_R}}
\end{figure*}

In Figure~\ref{fig:SAMI_MAGPI_BD_vs_R} we present the BD gradients for the SAMI and MAGPI galaxies, with the values listed in Tables~\ref{tab:bd_gradient_SAMI} and \ref{tab:bd_gradient_MAGPI}. It can be seen that negative gradients are apparent in both surveys across all mass bins and that the slopes become more negative with increasing stellar mass. Both the negative gradients and their mass-dependence are consistent with several previous studies at $z\lesssim0.1$ \citep[e.g.,][]{greener20, leeJC25, mailvaganam26a} and $1\lesssim z\lesssim 2$ \citep[e.g.,][]{nelson16a, tacchella18, matharu23, matharu25, ren26}. 
These trends are expected assuming galaxy centres contain more dense gas and dust (these quantities are linked via gas-to-dust vs gas-phase metallicity scaling relations; e.g., \citealt{deVis19}), which is typically observed for star forming galaxies at both low- \citep[e.g.,][]{casasola17} and high-redshifts \citep[e.g.,][and references therein]{hodge&daCunha20}. 

At a given stellar mass, the MAGPI galaxies appear slightly dustier than SAMI. This trend is expected if the dense gas fractions for galaxies in MAGPI are higher than SAMI and if the gas-phase metallicities of both samples are similar. Assuming the molecular-to-stellar mass ratio vs redshift scaling relations from \cite{tacconi18}, galaxies at a fixed stellar mass at $z=0.3$ have roughly twice as much dense gas as those at $z=0$. As shown in Figure~\ref{fig:MAGPI_MS_MZR}, MAGPI galaxies appear roughly consistent with the $z=0$ MZR relation, indicating that they will be comparable in metallicity. Assuming the MZR-redshift relations from \cite{sanders23}, the expected average metallicity evolution for galaxies over our mass range is $\sim$0.05~dex from $z=0.3$ to $z=0$. Taken together, these offer support for the trend that MAGPI galaxies contain more dust on average relative to SAMI galaxies.

\subsection{Balmer Decrement Dependence on Aperture Covering Fraction from SAMI and MAGPI}\label{BD_fc_compare}

The observed BD gradients will affect BD values measured from fixed aperture surveys (i.e., those relying on fibres or slits), which are commonly adopted for making global dust attenuation corrections. In the case of negative BD gradients, the BD derived from a smaller aperture ($f_c<1$) will always overestimate the true integrated BD value, becoming more pronounced as the covering fraction decreases. We now examine this effect in more detail. 
For the results of \textit{only} this Section, we restrict the SAMI sample to galaxies where the covering fraction reaches unity ($f_c=1$), which excludes 397 out of the 918 galaxies (43\%). This selection preferentially removes more massive sources. We infer $f_c$ as the ratio of stellar mass enclosed within an aperture relative to the total stellar mass as opposed to a fraction of flux from a specific band or emission line.

\begin{figure*}
$\begin{array}{cc}
	\includegraphics[width=0.46\textwidth]{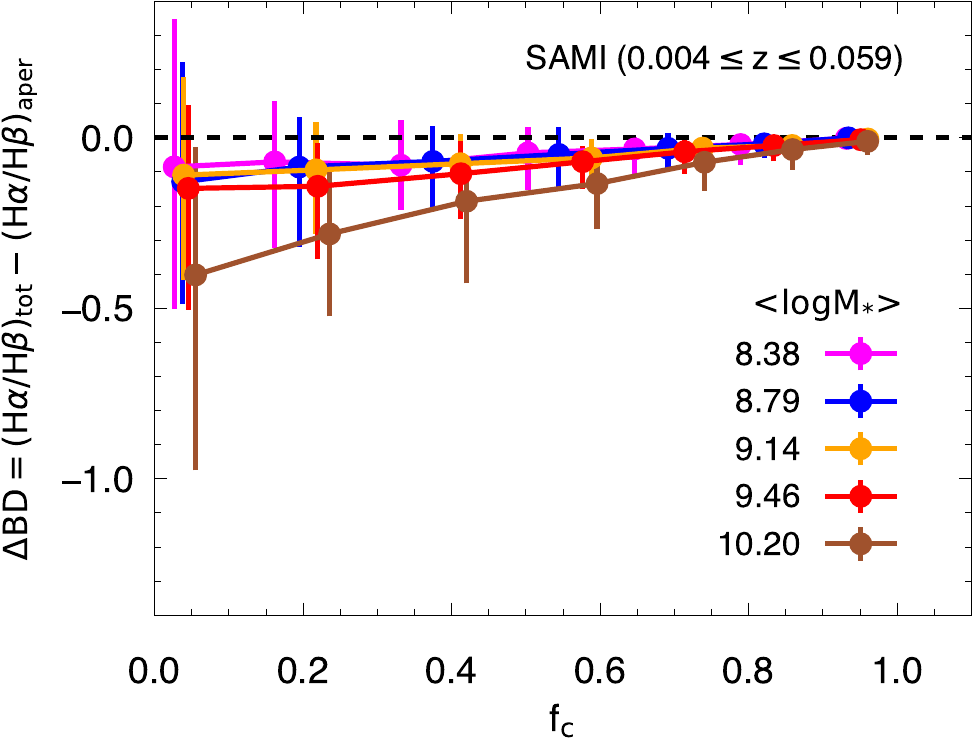} & \hspace{1mm}
\includegraphics[width=0.48\textwidth]{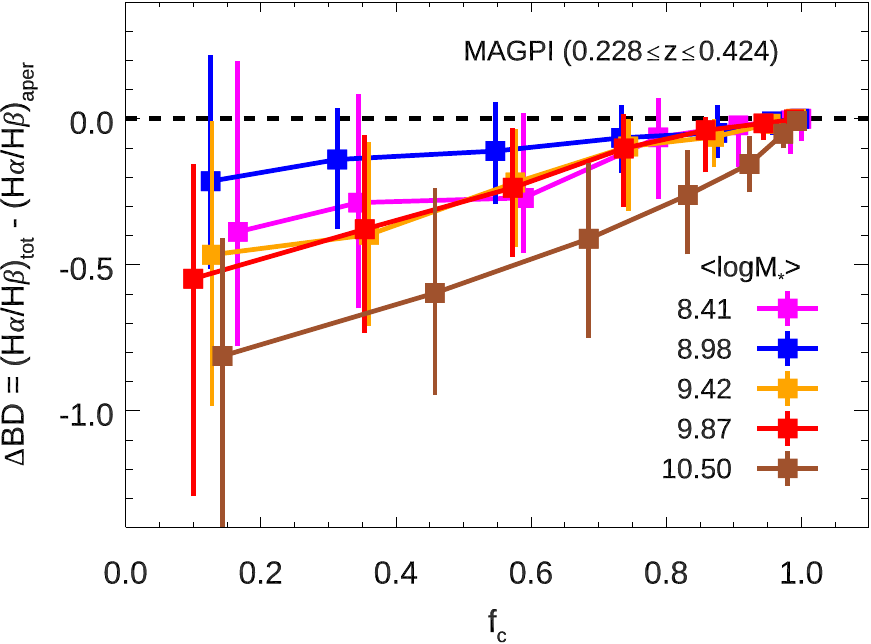} \\
\end{array}$
\vspace{-3mm}
    \caption{Difference in Balmer decrement ($\Delta \mathrm{BD}$) as a function of aperture covering fraction ($f_c$) in \textit{(Left:)} SAMI and \textit{(Right:)} MAGPI. The $\Delta \mathrm{BD}$ is based on the total aperture, (\halpha/\hbeta)$_\mathrm{tot}$ (i.e., $f_c=1$), and a smaller aperture, (\halpha/\hbeta)$_\mathrm{aper}$ (i.e., $f_c<1$), and $f_c$ is the fraction of stellar mass enclosed within an aperture. The data are separated into five mass-bins and subdivided into seven equal-number aperture bins. The errorbars indicate the 16th and 84th percentiles of the intrinsic scatter and not the measurement uncertainty. The trends are similar between samples (redshift), with more massive galaxies having more negative $\Delta \mathrm{BD}$ at lower $f_c$ values. The MAGPI $\Delta \mathrm{BD}$ are systematically more negative than the SAMI values at similar $f_c$ and mass. The scatter in $\Delta \mathrm{BD}$ values increases significantly at lower $f_c$, highlighting there is large uncertainty in using measurements from small $f_c$ to infer global BD values for individual sources. 
    \label{fig:SAMI_MAGPI_BDdiff_vs_fc}}
\end{figure*}

The difference in Balmer decrement from the total aperture (i.e., $f_c=1$) and a smaller aperture (i.e., $f_c<1$) as a function of $f_c$ is shown in Figure~\ref{fig:SAMI_MAGPI_BDdiff_vs_fc} and the values are provided in Table~\ref{tab:BDdiff_fc_SAMI} and \ref{tab:BDdiff_fc_MAGPI}. The tighter grouping of bins at large $f_c$ for more massive galaxies is due to more massive galaxies having steeper mass profiles such that a majority of mass is in the centre. In these cases, adding progressively larger apertures leads to smaller $f_c$ than for lower mass galaxies that tend to have flatter stellar mass profiles. There are clear trends for both samples that more massive galaxies have more negative $\Delta \mathrm{BD}$ at lower $f_c$ values, but the MAGPI $\Delta \mathrm{BD}$ are systematically more negative than the SAMI values at similar $f_c$ and mass. The scatter in the $\Delta \mathrm{BD}$ values increases significantly as $f_c$ decreases. This highlights that there is large uncertainty in using measurements from surveys with small $f_c$ to infer global BD values for individual sources. 

In theory, $\Delta \mathrm{BD}$-$f_c$ relations could be used to make average corrections for surveys that have $f_c<1$. However, these values depend significantly on the BD radial gradients at the respective redshift, which differ notably between the SAMI and MAGPI samples, and are poorly constrained at higher redshifts. Such application is further complicated by the fact that $f_c$ will be stellar mass-dependent at a fixed redshift and slit/fibre size, due to the size-mass relation of galaxies \citep[e.g.,][]{vanderWel14}. For these reasons, we do not attempt to make such corrections in subsequent Sections and caution against applying the SAMI and MAGPI $\Delta \mathrm{BD}$-$f_c$ values on higher redshift samples.

\subsection{Aperture Dependence of the Balmer Decrement-Stellar Mass Relation from SAMI and MAGPI}\label{BD_mstar_compare}
We now use the SAMI and MAGPI samples to demonstrate the importance of the spectroscopic aperture choice for the BD-$M_\star$ relationship. We consider two choices of apertures. The first is intended to match SDSS-III \citep[DR7;][]{abazajian09}, which used a circular fibre with a 3\arcsec\ diameter, and will be referred to as the `SDSS-like aperture'. For reference, star forming galaxies in SDSS-III have a median covering fraction of $f_c\sim0.25$ of the total flux or stellar mass \citep[e.g.,][]{zahid13}. The second aperture is intended to represent the total integrated flux (or very close to total in the case of SAMI) and will be referred to as the `integrated aperture'. Using the integrated aperture for both quantities provides the most robust manner for examining the true global BD-$M_\star$ relationship of galaxies.
We adopt the same criteria described in the previous section for restricting the spaxels that are included when summing fluxes within these apertures.  

For SAMI, which has a redshift range similar to SDSS-III, we adopt a 3\arcsec\ diameter circular aperture for the SDSS-like aperture and all spaxels inside the 15\arcsec\ hexabundle for the integrated aperture. 
For MAGPI, we adopt a 0.8\arcsec\ diameter circular aperture for the SDSS-like aperture and all spaxels inside the dilated mask for the integrated aperture. A 0.8\arcsec\ diameter at $z=0.3$ corresponds roughly the same physical size of 3.8~kpc that a 3\arcsec\ diameter aperture covers at $z=0.066$ (median redshift of our SDSS-III comparison sample).

\begin{figure*}
$\begin{array}{cc}
	\includegraphics[width=0.48\textwidth]{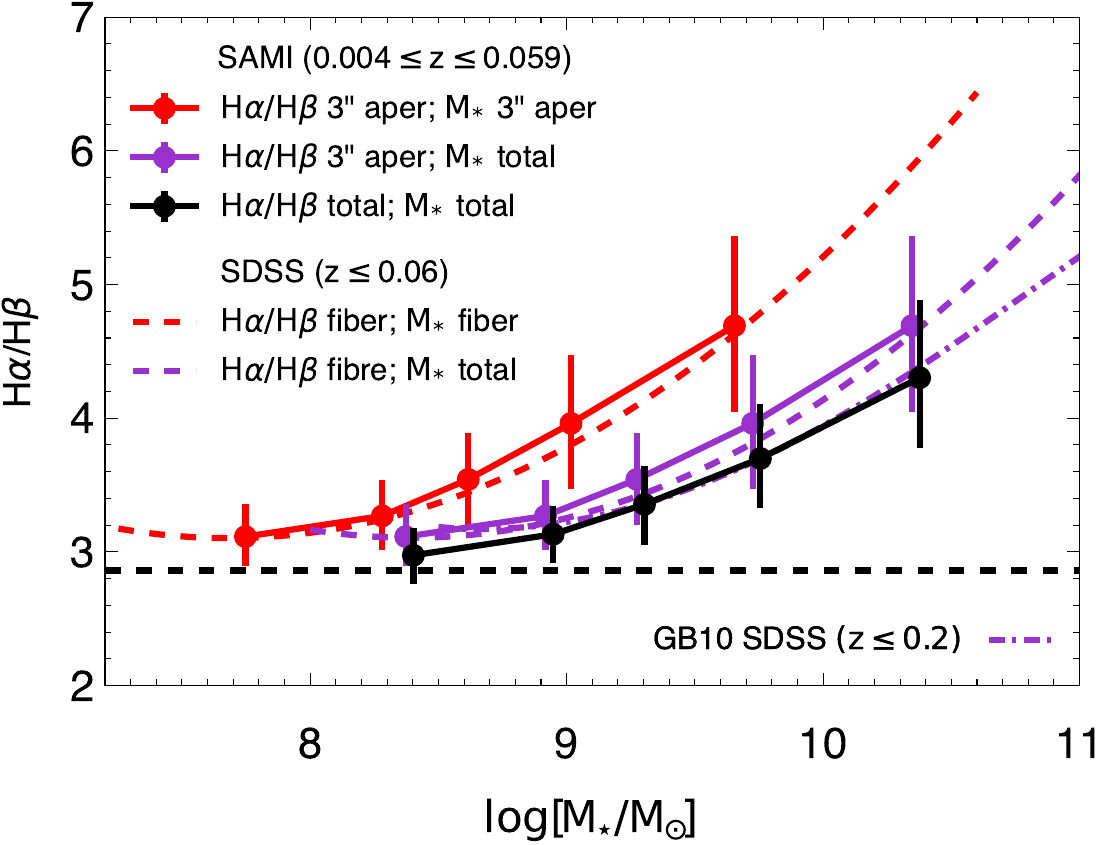} & \hspace{1mm}
\includegraphics[width=0.48\textwidth]{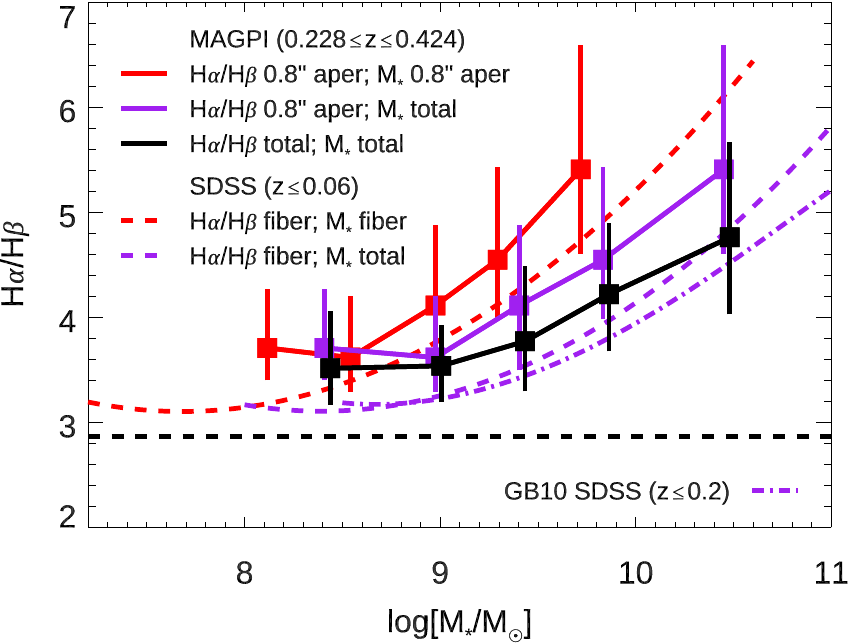} \\
\end{array}$
 \vspace{-3mm}
    \caption{Balmer decrement vs stellar mass ($M_\star$) for galaxies in \textit{(Left:)} SAMI and \textit{(Right:)} MAGPI separated into five mass-bins. The different colours denote different adopted apertures for the BD and $M_\star$ measurements, with a cartoon visualisation provided in Figure~\ref{fig:cartoon_BD_vs_M}. Purple and black points have same stellar mass, but are offset slightly for clarity. The offset between the BD of the purple and black values increases with increasing $M_\star$ and is a consequence of decreasing $f_c$ and steeper BD radial gradients at larger $M_\star$ (Figure~\ref{fig:SAMI_MAGPI_BD_vs_R}).  
    The errorbars indicate the 16th and 84th percentile of intrinsic scatter and not the measurement uncertainties, where the latter are usually smaller (e.g., Figure~\ref{fig:SAMI_MAGPI_BD_vs_param_individual}). More massive galaxies tend to be dustier, as found in previous studies (see text). Relationships using SDSS-III data restricted to the SAMI redshift range ($z\leq0.06$) for different apertures are shown as the dashed lines, which are in good agreement with the SAMI SDSS-like aperture results. The relation from \citet{garn&best10} using the full SDSS-III sample ($z\leq0.2$; uses total $M_\star$) is shown as the purple dash-dot line. 
    The MAGPI BD-$M_\star$ relationship is offset to higher BD at a given $M_\star$ than SAMI, which favours a redshift evolution. 
    \label{fig:SAMI_MAGPI_BD_vs_M}}
\end{figure*}

\begin{figure}
\begin{center}
	\includegraphics[width=0.44\textwidth]{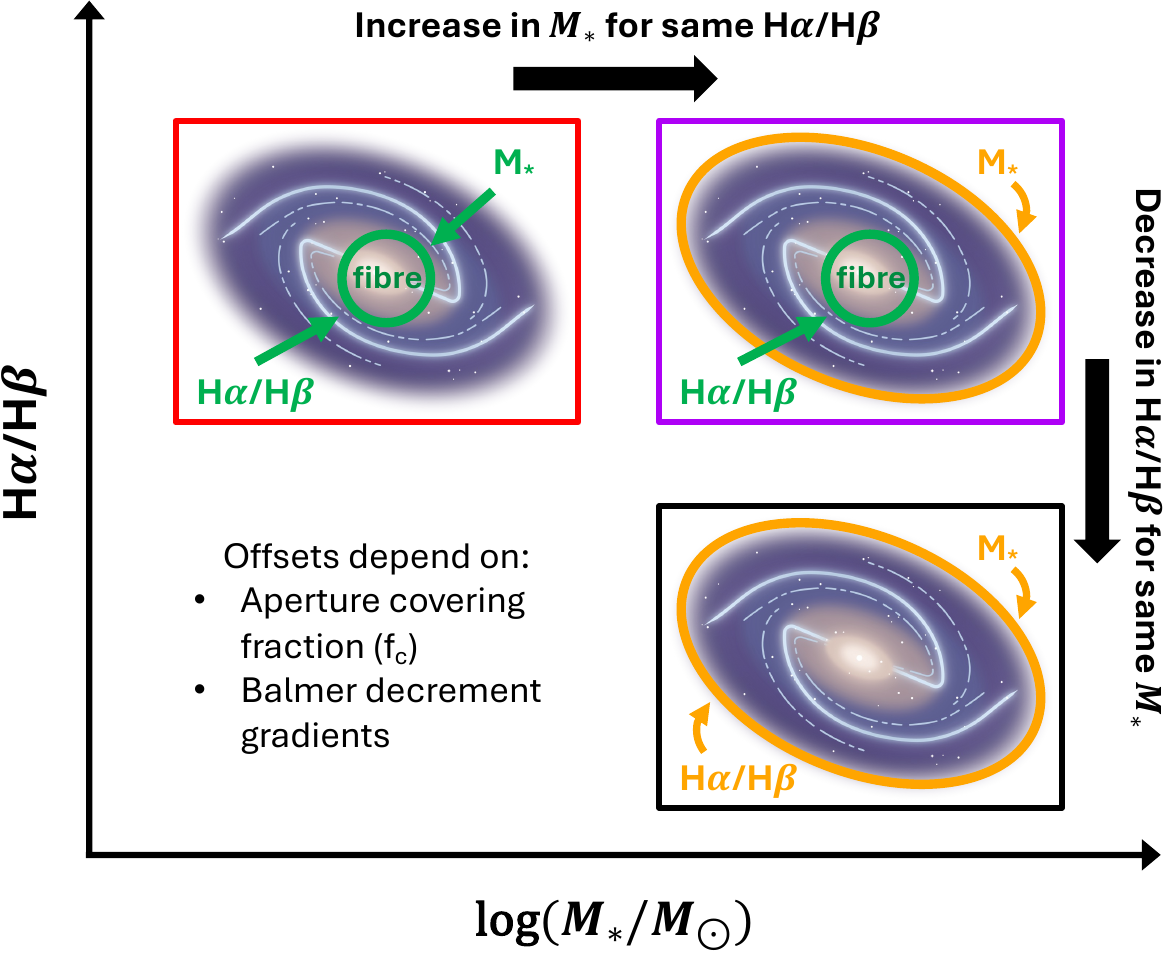}
 \end{center}
 \vspace{-1mm}
    \caption{Schematic representation of the offsets in the Balmer decrement vs stellar mass values shown in Figure~\ref{fig:SAMI_MAGPI_BD_vs_M}, with box outline colours matched to the corresponding symbol colours in that Figure. The green circle represents a fibre, with covering fraction, $f_c<1$, and the orange  ellipse is the full optical extent, with $f_c=1$. The $M_\star$ values will be larger with an increase in $f_c$ (red to orange box). Most star-forming galaxies tend to be dustier in their centre (i.e., negative BD gradient), resulting in higher average BD when using a fibre  relative to the entire galaxy (purple to black box). However, some galaxies may have no BD gradient or a positive BD gradient, resulting in no y-axis offset or a positive offset, respectively.  Measurements from SDSS-III, often adopted as a $z\sim0$ benchmark, usually correspond to the purple box. Cartoon galaxy image credit: NASA.
    \label{fig:cartoon_BD_vs_M}}
\end{figure}

In Figure~\ref{fig:SAMI_MAGPI_BD_vs_M}, we present the BD-$M_\star$ relationships for the SAMI and MAGPI galaxies when different apertures are adopted to measure each quantity. We provide the integrated aperture values in Tables~\ref{tab:bd_logM_SAMI} and \ref{tab:bd_logM_MAGPI}. A cartoon visualisation of the apertures and the typical offsets that occur is shown in Figure~\ref{fig:cartoon_BD_vs_M}. It can be seen in Figure~\ref{fig:SAMI_MAGPI_BD_vs_M}, that notable offsets occur between the different aperture choices, with the effects being most pronounced for more massive galaxies. The horizontal shift to larger masses (rightward) from using a SDSS-like aperture for both quantities (red points) to using a mixed aperture (SDSS-like aperture for BD and integrated aperture for $M_\star$; purple points) arises simply due to the stellar mass increasing with the larger aperture while the BD quantity remains the same. The size of this horizontal shift is larger for more massive galaxies because the covering fraction ($f_c$) of the SDSS-like aperture is smaller for more massive galaxies that are larger in physical size relative to lower-mass galaxies \citep[i.e., size-mass relation; e.g.,][]{vanderWel14}. There is a vertical shift to lower BD (downward) from using a mixed aperture (purple points) to an integrated aperture for both quantities (black points). These offsets are also larger for more massive galaxies because they have more negative (i.e., steeper) BD gradients relative to lower-mass galaxies (see Figure~\ref{fig:SAMI_MAGPI_BD_vs_R}), and therefore the BD values depend more strongly on the aperture size adopted. The nature of these shifts complicates our interpretation of a redshift evolution in the BD-$M_\star$ relationship when comparing between surveys with different $f_c$ values for the BD measurement and will be discussed in Section~\ref{discussion}. Comparing the integrated aperture BD-$M_\star$ relations from SAMI and MAGPI, there appears to be a higher BD at a given $M_\star$ for MAGPI relative to SAMI, indicating that galaxies at $z\sim0.3$ appear slightly dustier on average relative to $z\sim0$.

\begin{figure}
\begin{center}
	\includegraphics[width=0.44\textwidth]{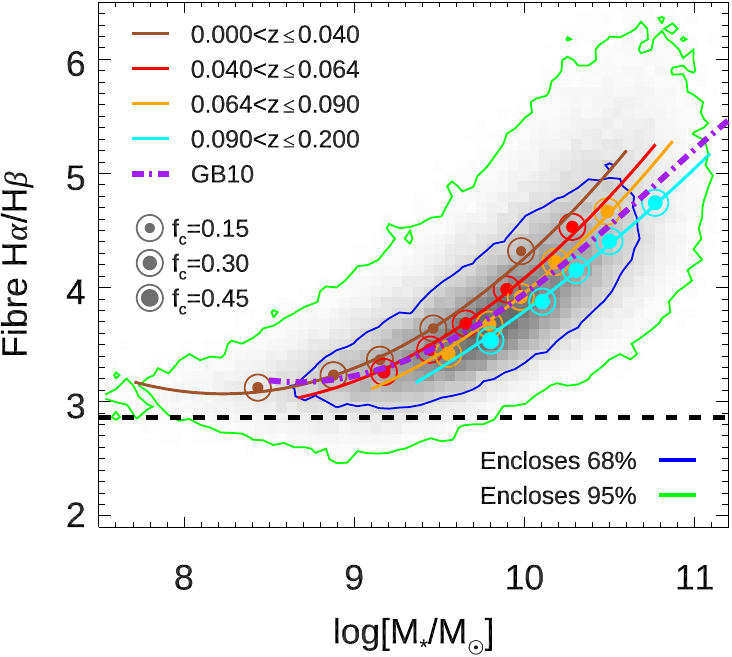}
 \end{center}
\vspace{-3.5mm}
    \caption{The grayscale 2D histogram shows the distribution of BD-$M_\star$ (BD from 3\arcsec\ fibre, total stellar mass) for star forming galaxies at $z\leq0.2$ in SDSS-III, which follows the \citet{garn&best10} relation. The coloured lines indicate the mean BD-$M_\star$ relations when subdividing into four equal-number redshift bins. The median covering fraction ($f_c$) in five mass bins for each redshift are denoted by the semi-filled symbols (see legend). At a fixed $M_\star$ ($\sim$fixed physical galaxy size), the covering fraction increases with increasing redshift due to galaxies having smaller angular size at higher redshift and a fixed fibre aperture size. Assuming a similar underlying `integrated aperture' BD-$M_\star$ relation for $z\leq0.2$ (which may not hold), this outcome is expected if galaxies have negative Balmer decrement gradients (see Figure~\ref{fig:cartoon_BD_vs_M}).  \label{fig:SDSS_BD_vs_M}}
\end{figure}

We also compare our results to the BD-$M_\star$ relationship using star forming galaxies in SDSS-III, which are restricted to residing in the SF-region of the BPT diagram (for more details of selection, see Section~2.4 of \citealt{battisti22}).
We find that the SDSS-like aperture for the SAMI data agrees very well with the SDSS-III relations when restricting it to the same redshift range ($z\leq0.06$), but that it differs from the \cite{garn&best10} relation that uses a wider redshift range ($z\leq0.2$). We attribute this difference primarily to the varying $f_c$ that the SDSS-III fibres have as a function of redshift and that the most massive galaxies preferentially reside at higher redshifts (due to requiring a larger volume to detect these rarer sources). We illustrate this effect in Figure~\ref{fig:SDSS_BD_vs_M} where we subdivide the SDSS-III sample by redshift. The median $f_c$ increases from 0.17, 0.24, 0.28, to 0.35 over the four increasing redshift bins shown. We attribute the observed offsets at a fixed $M_\star$, that the BD values are lower when $f_c$ is higher, as a consequence of most galaxies having a negative BD gradient as opposed to an actual underlying evolution across the $0.0\leq z\leq0.2$ range. Indeed, this is supported by the BD-$M_\star$ relation from the integrated aperture SAMI data aligning closest to the highest redshift SDSS-III subsample at the high-$M_\star$ end where the $f_c$ values are highest. At low-$M_\star$, galaxies show weaker gradients in their BD (Figure~\ref{fig:SAMI_MAGPI_BD_vs_R}) such that the SAMI and SDSS relations are expected to be similar regardless of $f_c$ (i.e., there are small offsets between purple and black points at lower $M_\star$ in Figure~\ref{fig:SAMI_MAGPI_BD_vs_M}). Interestingly, the integrated aperture SAMI relation (for $z\leq0.06$) remains relatively close to the \cite{garn&best10} line (for $z\leq0.2$), agreeing within the intrinsic scatter, although it is slightly lower at the low- and high-mass ends.

\subsection{Balmer Decrement Correlation Strength with Global Properties}\label{BD_correlation_compare}

The total stellar mass is one of the most reliably measured galaxy properties \citep[e.g.,][]{conroy13}. This is due to having a lower influence by the effects of dust attenuation (most of the mass is usually in lower-mass stars that dominate optical/near-IR) relative to many other properties and because of the ease that it can be measured observationally (requiring only a handful of photometric bands). It is these factors that make the use of total stellar mass as an indirect proxy of dust attenuation appealing. However, from a physical standpoint, it is not intuitive that a strong correlation between dust attenuation and total stellar mass should occur and that it could instead be a consequence of secondary correlations arising from the star-forming main sequence \citep[SFR-$M_\star$ relation; e.g.,][]{brinchmann04}, the Kennicutt-Schmidt relation \citep[$\Sigma_\mathrm{SFR}$-$\Sigma_\mathrm{mol}$;,e.g.,][]{kennicutt98} and/or gas-to-dust relation. 

The topic of primary and secondary correlations between the BD and galaxy parameters was explored in \cite{maheson24} using SDSS-III data, which found that stellar mass showed the strongest correlation relative to SFR, gas-phase metallicity, and gas velocity dispersion. However, given the role of aperture size on BD values (Section~\ref{BD_mstar_compare}), it is worthwhile to re-examine these correlations to determine if different results are found.

\begin{figure*}
\includegraphics[width=1\textwidth]{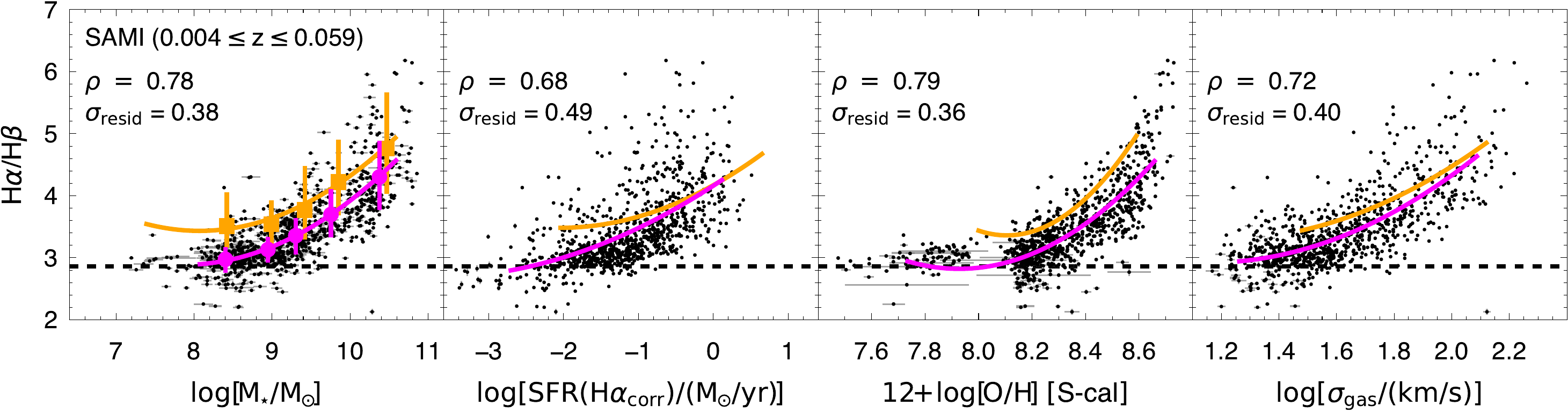} \\ \vspace{4mm}
\includegraphics[width=1\textwidth]{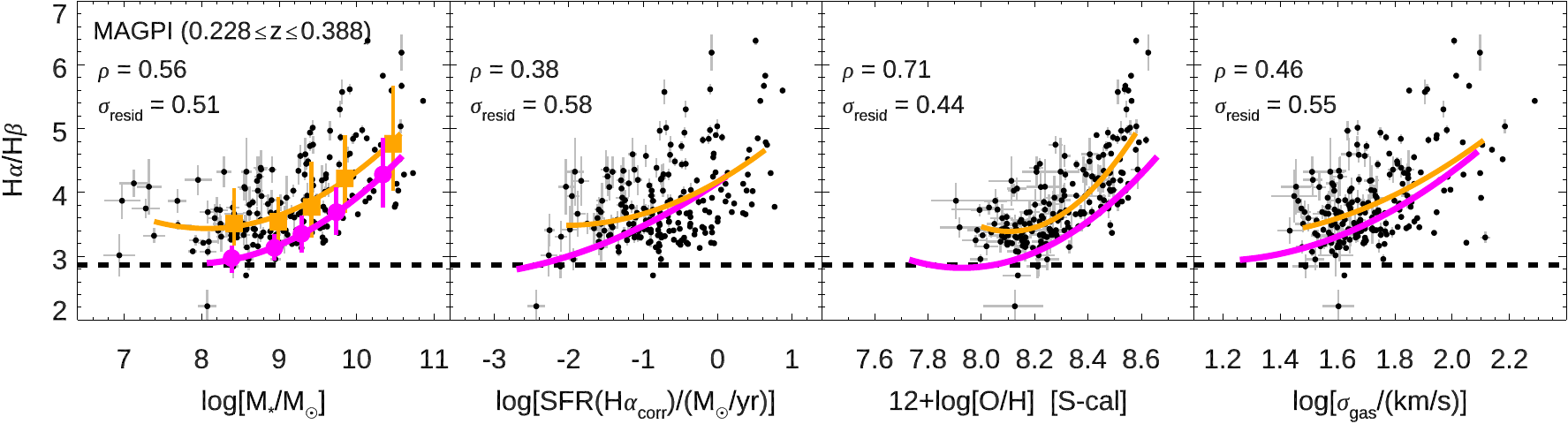}
 \vspace{-5mm}
    \caption{Integrated aperture Balmer decrement vs stellar mass ($M_\star$), SFR, gas-phase metallicity (12+log[O/H]), and gas velocity dispersion ($\sigma_\mathrm{gas}$) for galaxies in \textit{(Top:)} SAMI and \textit{(Bottom:)} MAGPI. The magenta and orange lines are second-order polynomial fits to the SAMI and MAGPI data (small black circles), respectively. Fit parameters are provided in Table~\ref{tab:bd_param_SAMI_MAGPI}. The large coloured symbols in the $M_\star$ panels correspond to the integrated-aperture bin values from Section~\ref{BD_mstar_compare}. The spearman correlation coefficients, $\rho$, are indicated in each panel with the high correlations found for stellar mass and gas-phase metallicity. The gap in the SAMI metallicity distribution arises due to the N2 inflection point in the $S$-calibration \citep{pilyugin&grebel16}. For both surveys, the strongest correlations are with $M_\star$ and metallicity. 
    \label{fig:SAMI_MAGPI_BD_vs_param_individual}}
\end{figure*}

We show the behaviour between the integrated-aperture BD and the $M_\star$, SFR, gas-phase metallicity, and gas velocity dispersion ($\sigma_\mathrm{gas}$) of SAMI and MAGPI in Figure~\ref{fig:SAMI_MAGPI_BD_vs_param_individual}. 
The SAMI data products do not provide an `integrated-aperture' $\sigma_\mathrm{gas}$, so we adopt the values for the $1R/R_e$ aperture. We use the same aperture for MAGPI and note that the $\sigma_\mathrm{gas}$ values are very similar between the $1R/R_e$ aperture and the undilated/dilated mask apertures. 
Interestingly, we find that the spearman rank correlation coefficients, $\rho$, for the metallicity are similar to the stellar metallicity and that these are higher than SFR and $\sigma_\mathrm{gas}$. The importance of metallicity contrasts with the result of \cite{maheson24}. This may be due to the role of aperture size, since galaxies can also show strong metallicity gradients \citep[e.g.,][]{maiolino&mannucci19}, but it may also be attributed to their choice to rely on a combination of different metallicity diagnostics because their values can differ significantly from each other \citep[e.g.,][]{kewley&ellison08}. We choose not to perform a detailed comparison of correlations with different metallicity diagnostics in this work, since metallicity is not the primary focus. Similarly to \cite{maheson24}, we find a moderate correlation between BD and $\sigma_\mathrm{gas}$. \cite{maheson24} suggest this could be driven by $\sigma_\mathrm{gas}$ tracing the gravitational potential of galaxies (linked to stellar mass), and that more massive galaxies are better able to retain their metals (linked to metallicity) from feedback processes relative to lower-mass galaxies. 

As highlighted in \cite{maheson24}, the SFR derived from dust-corrected \halpha\ is not independent of the Balmer decrement since it is used directly in the colour excess term (see Eq.~\ref{eq:EBVgas}). Those authors argue in favour of using SFRs derived from SED fitting for comparison instead. However, these trace different SFR timescales ($\sim$5Myr for \halpha\ vs $\sim$100Myr for SED) that complicate this interchange. 
We note the correlation of these parameters as a caveat, but is it interesting that despite this the SFR(\halpha) correlations are weaker than those of stellar mass and metallicity. 

Our take-away from this comparison is that stellar mass shows a strong correlation with the BD with respect to other galaxy properties when using integrated aperture values, which reinforces the assertion that it is a reasonable proxy for the amount of nebular attenuation in galaxies. This is beneficial because stellar mass is easier to measure observationally than the other properties and less reliant on circularity issues from dust corrections (required for SFR(\halpha) and metallicities based on $S$-cal). A more detailed comparison of the spatially-resolved correlation between the BD and other galaxy properties in MAGPI is presented in \citet{mailvaganam26b}.

\begin{table}
\caption{Fit parameters, $p_x$, and spearman correlation coefficient, $\rho$, for integrated aperture Balmer decrements against stellar mass, SFR, gas-phase metallicity, and gas velocity dispersion for SAMI and MAGPI galaxies. Fits have the form $\mathrm{BD} = p_0 + p_1x+p_2x^2$ \label{tab:bd_param_SAMI_MAGPI}} 
\begin{center}
\begin{tabular}{cccccc}
 \hline \\[-1em]
 \multicolumn{6}{c}{SAMI ($0.004 < z< 0.059$)} \\
 $x$ & range & $p_0$ & $p_1$ & $p_2$ & $\rho$ \\
\hline
 log$M_\star$ & [8.07, 10.59] & 16.934 & -3.5719 & 0.22705 & 0.78 \\
 logSFR & [-2.70, 0.12] & 4.1612 & 0.786 & 0.10407 & 0.68 \\
 12+log[O/H] & [7.73, 8.66] & 207.41 & -51.616 & 3.2555 & 0.79 \\
 $\log\sigma_{\mathrm{gas}}$ & [1.26, 2.09] & 5.5477 & -4.5517 & 1.9713 & 0.72 \\
\hline \\

\hline \\[-1em]
 \multicolumn{6}{c}{MAGPI ($0.228 < z< 0.388$)} \\
 $x$ & range & $p_0$ & $p_1$ & $p_2$ & $\rho$ \\
\hline
 log$M_\star$ & [7.39,10.58] & 18.923 & -3.8389 & 0.23786 & 0.56 \\ 
 logSFR & [-2.04, 0.65] &  4.1687 & 0.66906 & 0.16305 & 0.38 \\ 
 12+log[O/H] & [8.00, 8.59] & 452.41 & -110.80 &  6.8348 & 0.71 \\ 
 log$\sigma_\mathrm{gas}$ & [1.48, 2.12] & 5.8623 & -4.3124 &  1.8089 & 0.46 \\ 
\hline
 \end{tabular}
\end{center}
\end{table}

\section{Discussion - Redshift evolution of the Balmer Decrement-Stellar Mass Relation}\label{discussion}

As noted in Section~\ref{intro}, several studies have examined the BD-$M_\star$ relation and made comparisons between surveys probing different epochs. Most of these studies find relationships that are roughly consistent with the local relation of \cite{garn&best10} based on SDSS-III, which, fortuitously, is not very different from the integrated aperture relation from SAMI for $z\leq0.06$. However, as shown in Section~\ref{BD_mstar_compare}, the value of $f_c$ has a noticeable impact on the SAMI and MAGPI galaxies due to the negative BD gradients. To mitigate for this effect, we restrict our comparison to studies in which the spectroscopic data have $\left<f_c\right> \gtrsim 0.5$. We also restrict it to surveys with $\gtrsim$200 galaxies in their sample and that they are representative of normal star-forming galaxies at their redshift (based on residing roughly on the galaxy main sequence). A caveat of comparing samples across redshift is that they have different selection functions and/or line-detection requirements. In particular, requiring the detection of the BD can bias samples toward less-dusty sources.

\begin{figure*}
$\begin{array}{lr}
	\includegraphics[width=0.48\textwidth]{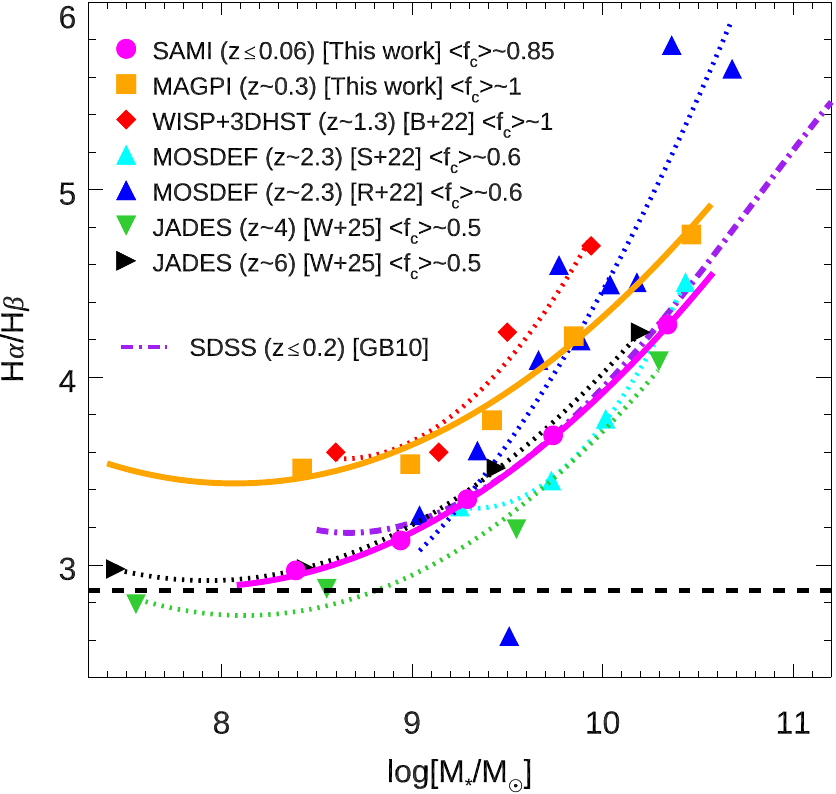} & \hspace{1mm}
\includegraphics[width=0.47\textwidth]{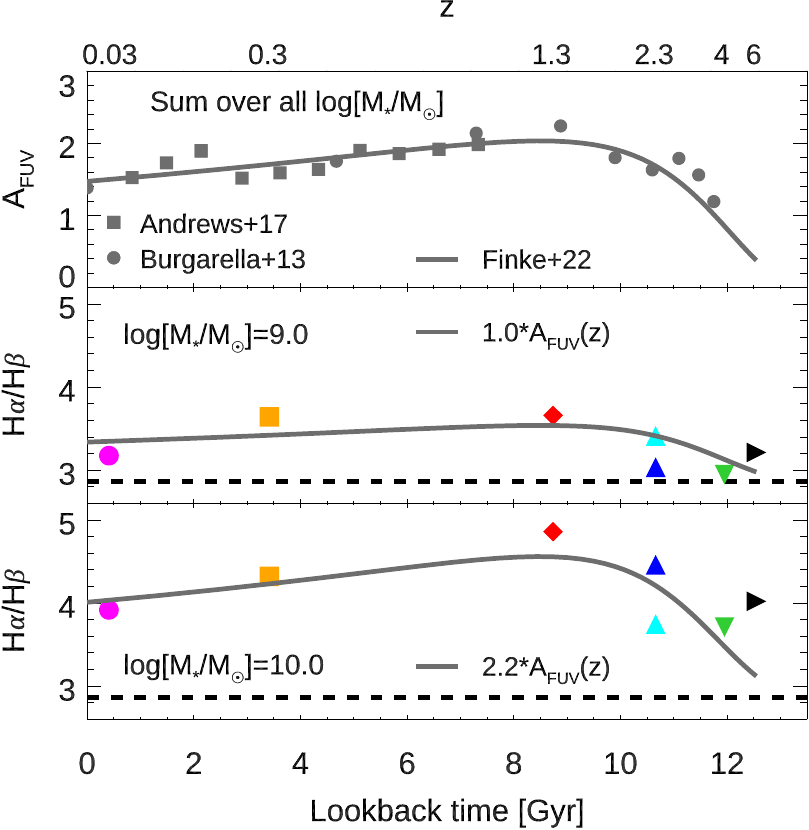} \\
\end{array}$
 \vspace{-1mm}
    \caption{\textit{(Left:)} BD-$M_\star$ relation from SAMI and MAGPI compared to the literature \citep{battisti22, shapley22, runco22, woodrum25}. Coloured lines correspond to second-order polynomial fits to the data (line colours match symbols) and are used to determine a representative value at fixed $M_\star$ for the right panel. \textit{(Right:)} The top panel shows the $A_\mathrm{FUV}$-$z$ relations from observations \citep{burgarella13, andrews17b} and the model of \citet{finke22} based on the extragalactic background light. The two lower panels show the BD-$z$ relation (in terms of lookback time) using values at log$[M_\star/M_\odot]=9$ and 10 from the fits in the left panel. These are qualitatively consistent with rescaled versions of the \citet{finke22} relation (see Eq.~\ref{eq:Afuv_BD}; scaling $f$ values indicated in legends), which are not fits and shown only for guidance of redshift evolution trends.}
    \label{fig:MAGPI_BD_vs_M_compare}
\end{figure*}

We briefly summarise the samples used in our comparison.
\cite{battisti22} used stacked observations of $\sim$900 \halpha-detected galaxies in the WISP \citep{atek10, battisti24} and 3D-HST \citep{brammer12} surveys, which used the HST/WFC3 grism and can be considered as having $\left<f_c\right>\sim1$ due to the slitless nature of the observations. 
\cite{shapley22} and \cite{runco22} use $\sim$200 and $\sim$500 galaxies, respectively, in the MOSDEF survey \cite{kriek15}, which used the Keck/MOSFIRE multi-object slit spectrograph \citep[0.7\arcsec\ slits;][]{mcLean12} that provides an average light-weighted covering fraction of $\left<f_c\right>\sim0.63$ for the sample (priv. comm.). There are noticeably different BD-$M_\star$ trends recovered between \cite{shapley22} and \cite{runco22}, which may be a consequence of the former relying on BD-detections in individual sources that may bias it toward less-dusty sources whereas the latter relies on stacking [NII]-detected sources that retain a larger fraction of the MOSDEF parent sample. 
\cite{woodrum25} used $\sim$600 galaxies with individually detected BD in the JADES survey \citep{dEugenio25}, which used the JWST/NIRSpec micro-shutter array that provided an average $\left<f_c\right>\sim0.5$ for the sample (priv. comm.).

In Figure~\ref{fig:MAGPI_BD_vs_M_compare}, \textit{Left}, we compare the BD-$M_\star$ relation from SAMI and MAGPI to several studies at higher redshifts with $\left<f_c\right> \gtrsim 0.5$.
We use second-order polynomial fits to each dataset and overlay these for guidance of trends, with the fit parameters for the SAMI and MAGPI values taken from Table~\ref{tab:bd_param_SAMI_MAGPI}. 
It is apparent that a redshift evolution is preferred, which is easier to see in the two lower panels in Figure~\ref{fig:MAGPI_BD_vs_M_compare}, \textit{Right}, where we show BD values inferred at $\log [M_\star/M_\odot]=9$ and 10 (based on polynomial fits) as a function of redshift. Taking into account the lower $\left<f_c\right>$ of the SAMI, MOSDEF, and JADES samples, their BD values could be lower for the highest $M_\star$ bins if negative Balmer decrement gradients are present (not currently well constrained at high-$z$; see Section~\ref{BD_fc_compare}). This would act to make the differences with MAGPI and WISP+3DHST even larger, and add further support for a redshift evolution.

Several observational studies combining UV-to-IR photometry have derived the average UV (stellar continuum) dust attenuation, $A_\mathrm{UV}$, (or similarly the obscuration fraction $L_\mathrm{UV}/L_\mathrm{IR}$) of galaxies as a function of redshift \citep[e.g.,][]{burgarella13, andrews17b, zavala21}. These generally find that $A_\mathrm{UV}$ increases from the local Universe ($z\sim0$) to cosmic noon ($1\lesssim z\lesssim2$) and then decreases toward higher redshifts. Similar findings result from the use of cosmic extragalactic background light \citep[EBL; e.g.,][]{andrews18, finke22, chiang25}. We provide some of the $A_\mathrm{FUV}$-$z$ results from the literature in the upper panel of Figure~\ref{fig:MAGPI_BD_vs_M_compare}, \textit{Right}, for reference. However, there are also studies suggesting little redshift evolution in the obscuration fraction of galaxies at fixed stellar mass over $0\lesssim z \lesssim2$ \citep[e.g.,][]{whitaker17, shivaei24} that appear to be in tension with these results \citep[cf.][]{vanderWel25}. 

An evolving BD-$M_\star$ relation with redshift is expected under a simplistic assumption that the attenuation of the ionised gas scales linearly with the amount of stellar attenuation (i.e., $A_{\mathrm{H}\alpha} \propto A_\mathrm{UV}$), but noting that linking the reddening of these components is non-trivial due to the fact that they trace different regions and stellar population ages. This correlation holds in galaxies at low-$z$ \citep[e.g.,][]{calzetti94, kreckel13, battisti16, zahid17, emsellem22, battisti25a} and at cosmic noon \citep[$1\lesssim z \lesssim 2$; e.g.,][]{reddy15, puglisi16, reddy20}. The correlation appears weaker at $z\gtrsim4$ \citep[e.g.,][]{woodrum25, tsujita26, karthikeyan26}, although the dust attenuation values at these redshifts are very low, which makes these quantities challenging to measure. 

Assuming a linear scaling in the attenuation between ionised gas and stellar attenuation, we can relate $A_\mathrm{FUV}$ to a Balmer decrement through
\begin{equation}
    \frac{F(\mathrm{H}\alpha)}{F(\mathrm{H}\beta)}\simeq 2.863\times10^{(f A_\mathrm{FUV}\,0.4 (k(\mathrm{H}\beta)-k(\mathrm{H}\alpha))/ k(\mathrm{FUV})) } \,,
	\label{eq:Afuv_BD}
\end{equation}
where $f$ is a constant that accounts for differential reddening between the stellar continuum and ionised gas, and we use the stellar continuum dust attenuation curve of \cite{calzetti00}, with $k(\mathrm{FUV})=10.27$. We note that both the $f$ and $k(\mathrm{FUV})$ values may vary with stellar mass and/or redshift, however, our intention here is to use this relation only for visual guidance and not to fit the data. 
As seen in the lower panels of Figure~\ref{fig:MAGPI_BD_vs_M_compare}, \textit{Right}, the functional behaviour of BD-$z$ at fixed $M_\star$ appears roughly consistent with $A_\mathrm{FUV}$-$z$ (integrated over all $M_\star$).

A notable aspect to the $A_\mathrm{UV}$-$z$ relation is that the redshift at which the attenuation is similar to the local Universe is at $z\sim2.5$, which is close to the mean of the MOSDEF survey. This could explain the lack of evolution suggested in \cite{shapley22}, without the need for changes in the dust mass absorption coefficient, dust-to-gas mass ratio, and/or dust distribution (see their Section~4), although those factors may indeed vary. A detailed characterisation of these quantities could help to understand the nature of the evolution of BD-$M_\star$ but this is beyond the scope of this paper. Finally, we also note that the BD-metallicity relation shows a slightly higher correlation than that of BD-$M_\star$ for MAGPI, presumably because metals are fundamentally related to dust formation \citep{galliano18}. The metallicity evolution of galaxies also likely plays an important role in why BD values (dust attenuation) at high-$z$ will be lower due to lower metallicities despite galaxies having high gas fractions.

\section{Conclusions}\label{conclusion}
We present the MAGPI survey emission line products data release, delivered through the Data Central archive, which includes resolved emission line maps for 836 galaxies at $0.05\leq z_\mathrm{spec}\leq 0.424$ (\halpha\ window) and aperture-based emission line measurements for 2,607 galaxies at $0.05\leq z_\mathrm{spec}\leq 1.5$, for all 56 MAGPI fields. The primary goal of the MAGPI survey is to study the physical drivers of galaxy transformation at a lookback time of 3–4 Gyr ($0.26\leq z\leq 0.42$), and therefore the largest fraction of the galaxy sample resides at $z\sim0.3$, however, there are thousands of additional galaxies serendipitously detected in the MUSE field-of-view that can be studied and are included as part of this release. 

We use our data products to examine the role that spectroscopic aperture covering fraction ($f_c$) has on the measurements of the BD, finding that they can be substantial at $f_c\lesssim0.5$ because galaxies typically have negative BD radial gradients at low-$z$, which are more negative for more massive galaxies. This results in a systematic decrease in the BD values that are derived using integrated apertures relative to fixed apertures that cover a lower fraction of the galaxy extent. Accounting for this effect is critical when examining the BD-$M_\star$ relation between surveys that have different $f_c$ due to their fibre or slit sizes. It also implies that dust corrections on emission lines measured in fibre or slit programs may be overestimated if the dust corrections are based on BD values. 

Comparing spectroscopic surveys with $\left<f_c\right> \gtrsim 0.5$, we find that the BD-$M_\star$ relation varies with redshift and is roughly consistent with the behaviour of UV stellar continuum attenuation with redshift ($A_\mathrm{FUV}$-$z$), peaking at $z\sim 1.2$. If true, this behaviour would imply that for a fixed stellar mass, the average dust attenuation of ionised gas in galaxies is largest at $z\sim 1.2$ and lower at both lower and higher-redshifts. Our assertion of a redshift evolution in the BD-$M_\star$ relation hinges on limited samples at intermediate redshifts ($0.2\lesssim z \lesssim2$) that cover $\sim$60\% of the age of the universe. It also does not account for evidence suggesting that galaxies at higher redshifts ($1.5\lesssim z \lesssim4.5$) may be optically thick to Balmer emission lines \citep[e.g.,][]{reddy26}, leading to an underestimation of the true amount of dust attenuation. 
Characterising this relation at $1\lesssim z\lesssim2$ with large samples from JWST, Euclid, and Roman grism data will be critical to verify its behaviour. Massively multiplexed fibre-based surveys like DESI \citep{DESI26}, various 4MOST programs (e.g., 4HS \citealt{taylor23} and WAVES \citealt{driver19}), and Subaru/PFS \citep{greene22} could also be useful to fill the $0.2\lesssim z \lesssim1$ window; however, it will be important to account for the impact of covering fractions on the Balmer decrement measurements. 

Lastly, we highlight that the SAMI and MAGPI surveys probe a range of galaxy environments, but this aspect was not explored in this work. In a future paper, we plan to explore the role of galaxy environment on the amount of dust attenuation \citep[e.g.,][]{liu25} in galaxies and on the shape of dust attenuation curves by combining with HST data (PI: Battisti) for 11 of the MAGPI fields (Battisti et al. in prep.).

\section*{Acknowledgements}
The authors thank the anonymous referee, whose suggestions helped to clarify and improve the content of this work. 
We thank the ESO staff, and in particular the staff at Paranal Observatory, for carrying out the MAGPI observations. MAGPI targets were selected from GAMA. GAMA is a joint European-Australasian project based around a spectroscopic campaign using the Anglo-Australian Telescope. GAMA was funded by the STFC (UK), the ARC (Australia), the AAO, and the participating institutions. GAMA photometry is based on observations made with ESO Telescopes at the La Silla Paranal Observatory under programme ID 179.A-2004, ID 177.A-3016. 
The MAGPI team acknowledges support by the Australian Research Council Centre of Excellence for All Sky Astrophysics in 3 Dimensions (ASTRO 3D), through project number CE170100013.
AJB thanks Mariska Kriek and Naveen Reddy for providing estimates of the covering fraction of the MOSDEF survey, Andy Bunker for providing details on the covering fraction of the JADES survey, and Gabriel Maheson for providing details on the Blue Jay survey. 
CF is the recipient of an Australian Research Council Future Fellowship (project number FT210100168) funded by the Australian Government. 
AFM acknowledges support from RYC2021-031099-I and PID2024-162088NB-I00 of MICIN/AEI. 
EG acknowledges the support of the National Natural Science Foundation of China (NSFC) under grants NOs. 1251101411, 12533003, 1257030642. 
EG acknowledges the Program for Innovative Talents, Entrepreneur in Jiangsu. 
TG acknowledges support from the Australian Research Council through Discovery Project DP210101945, funded by the Australian Government. 
SMS acknowledges funding from the Australian Research Council (DE220100003).
LMV acknowledges support by the German Academic Scholarship Foundation (Studienstiftung des deutschen Volkes) and the Marianne-Plehn-Program of the Elite Network of Bavaria.
We acknowledge the invaluable labour of the maintenance and clerical staff at our institutions, whose contributions make our scientific discoveries a reality. 
This research was conducted on the indigenous land of the Ngunnawal and Whadjuk Noongar people. 

\section*{Data Availability}
All MAGPI VLT/MUSE spectroscopic data used in this paper are publicly available through the ESO archive science portal\footnote{\url{https://archive.eso.org/scienceportal/home}} (ESO program ID: 1104.B-0536). This data release adds the following data products to the Data Central website for MAGPI\footnote{\url{https://docs.datacentral.org.au/magpi/}}: (1) emission line maps for individual galaxies at $z\leq 0.424$, (2) aperture-based integrated emission line measurements for individual galaxies at $z\leq 1.50$, (3) integrated SFR measurements for individual galaxies at $z\leq 1.50$. These products are also available in the Supplementary Data material, however, we note the latter are static and not subject to version updates. 
Other data products can be made available upon reasonable request to the first author and/or by joining the MAGPI survey team (contact details available at \url{http://magpisurvey.org}).



\bibliographystyle{mnras}
\bibliography{AJB_bib}


\section*{Affiliations}
\noindent
$^{1}$International Centre for Radio Astronomy Research, University of Western Australia, 7 Fairway, Crawley, WA 6009, Australia \\
$^{2}$Research School of Astronomy and Astrophysics, Australian National University, Cotter Road, Weston Creek, ACT 2611, Australia\\
$^{3}$ARC Centre of Excellence for All Sky Astrophysics in 3 Dimensions (ASTRO 3D), Australia\\
$^{4}$School of Physics, University of New South Wales, Kensington, NSW 2032, Australia\\
$^{5}$Department of Physics, Institute for Computational Cosmology, Durham University, South Road, Durham DH13LE, UK\\
$^{6}$Department of Physics, Centre for Extragalactic Astronomy, Durham University, South Road, Durham DH13LE, UK\\
$^{7}$School of Mathematical and Physical Sciences, Macquarie University, NSW 2109, Australia\\
$^{8}$Astrophysics and Space Technologies Research Centre, Macquarie University, Sydney, NSW 2109, Australia\\
$^{9}$Sydney Institute for Astronomy (SIfA), School of Physics, The University of Sydney, NSW 2006, Australia\\
$^{10}$Instituto de Astrof\'{i}sica e Ci\^{e}ncias do Espaço-Centro de Astrof\'isica da Universidade do Porto (IA-CAUP), Rua das Estrelas, 4150-762 Porto, Portugal\\
$^{11}$Department of Astronomy, University of Massachusetts-Amherst, Amherst, MA 01003, USA\\
$^{12}$School of Mathematics and Physics, University of Queensland, St Lucia, Queensland 4072, Australia\\
$^{13}$Instituto de Astrofísica de Canarias, Calle Vía Láctea, E-38206 La Laguna, Spain\\
$^{14}$Departamento de Astrofísica, Universidad de La Laguna, E-38206 La Laguna, Spain\\
$^{15}$Centre for Astrophysics and Supercomputing, Swinburne University, John Street, Hawthorn VIC 3122, Australia\\
$^{16}$School of Astronomy and Space Science, Nanjing University, Nanjing 210023, People’s Republic of China\\
$^{17}$Key Laboratory of Modern Astronomy and Astrophysics (Nanjing University), Ministry of Education, Nanjing 210023, People’s Republic of China\\
$^{18}$Australian Astronomical Optics, Macquarie University, Sydney, NSW 2109, Australia\\
$^{19}$Institut d'Astrophysique de Paris, CNRS, Sorbonne Universit\'e, 98bis Boulevard Arago, 75014, Paris, France\\
$^{20}$University of Strasbourg, CNRS UMR 7550, Observatoire astronomique de Strasbourg, F-67000 Strasbourg, France\\
$^{21}$SISSA– International School for Advanced Studies, Via Bonomea 265, I-34136 Trieste, Italy\\
$^{22}$UWC– University of the Western Cape, Department of Physics and Astronomy, Cape Town 7535, South Africa\\
$^{23}$IFPU– Institute for Fundamental Physics of the Universe, Via Beirut, 2, 34151 Trieste, Italy\\
$^{24}$INFN-Sezione di Trieste, via Valerio 2, I-34127 Trieste, Italy\\
$^{25}$Department of Astrophysics, University of Vienna, T\"urkenschanzstra{\ss}e 17, 1180 Vienna, Austria\\
$^{26}$Universitäts-Sternwarte, Fakultät für Physik, Ludwig-Maximilians-Universität München, Scheinerstr. 1, 81679 München, Germany


\appendix

\section{\gist\ configuration settings}\label{gist_config}

For the stellar continuum measurements, we use spectral templates from the C3K library (C. Conroy, priv. comm.), which are based on the Modules for Experiments in Stellar Astrophysics (\texttt{MESA}) Isochrones and Stellar Tracks (\texttt{MIST}; \citealt{choi16}). These spectral templates cover the wavelength range of $1500\leq \lambda_\mathrm{rest} \leq 8500$\AA.
We normalise the templates in a light-weighted manner (\texttt{NORM\_TEMP}). Before fitting, the templates are spectrally convolved to match the wavelength-dependent MUSE spectral resolution derived from skylines (e.g., \citealt{bacon17}), which are determined field-by-field. 
We adopt a multiplicative Legendre polynomial (\texttt{MDEG}) of order 12 for the stellar continuum fitting for sources that can reach our desired $S/N$ for Voronoi binning (details below); otherwise we adopt \texttt{MDEG}=4. 
During the stellar continuum fitting stage, the positions of common sky lines, the GLAO sodium laser region, and galaxy emission lines are excluded.

For the Voronoi binning of the stellar continuum, this is based on considering individual spaxels with $S/N>2$ (\texttt{MIN\_SNR}) over a specified wavelength range that reside inside the \profound\ dilated mask (see Section~\ref{MAGPI_data}; \texttt{DILATED\_MASK}=0) and excludes regions affected by skylines and emission/absorption lines of each galaxy. For galaxies at $z< 0.51$, we use the spectrum from $6050\leq \lambda_\mathrm{obs} \leq 7750$\AA, which corresponds to the peak sensitivity range of MUSE\footnote{\url{https://www.eso.org/sci/facilities/paranal/instruments/muse/inst.html}}. However, at $z\geq 0.51$ this observer-frame wavelength range begins probing $\lambda_\mathrm{rest} \leq 4000$\AA, which can be intrinsically faint in passive galaxies relative to $\lambda_\mathrm{rest} > 4000$\AA, due to absorption features and the lack of young, UV-bright stars (i.e., the 4000\AA\ break feature). For consistency in probing rest-frame optical wavelengths as much as possible, we adopt a sliding wavelength range of $4000\leq \lambda_\mathrm{rest} \leq 5132$\AA\ for $0.51<z\leq 0.82$ (this corresponds to $6050\leq \lambda_\mathrm{obs} \leq 7750$\AA\ at $z=0.51$). In other words, we adopt progressively redder regions of the MUSE spectrum for the estimation of $S/N$ with increasing redshift, up to $z=0.82$. At $z=0.82$, this window hits the red-end of the MUSE instrument (9351\AA) and we begin to use regions of the spectrum below rest-frame 4000\AA, adopting a fixed observer-frame window of $7288\leq \lambda_\mathrm{obs} \leq 9351$\AA\ (i.e., the same spectral window at $z=0.82$ is used for $z>0.82$).

We adopt two $S/N$ thresholds for the Voronoi binning, depending on the quality of the data. We first attempt $S/N=10$ and adopt this threshold if the galaxy has two or more bins that achieve this threshold. We also impose a restriction in the parameter range for stellar velocities and velocity dispersion of $-400\leq v_\mathrm{stellar}\leq400$~km/s and $1\leq \sigma_\mathrm{stellar}\leq500$~km/s, respectively, where $v_\mathrm{stellar}=0$ corresponds to the assigned redshift (from a 1\arcsec radius aperture). There are 554 galaxies (21.2\%) with sufficient continuum signal to create two or more Voronoi bins reaching the $S/N\gtrsim10$ threshold (the algorithm can also assign bins below the threshold in the outskirts). We note that the default \gist\ code does not impose a restriction on the stellar velocity range, but that its primary intended use is data with moderate to high continuum $S/N$ where such restrictions are not required. 
For sources with low continuum $S/N$ that do not achieve $S/N=10$ in multiple bins, we rerun \gist\ setting a $S/N=4$ threshold for the continuum binning. We impose a stronger restriction on stellar velocities and velocity dispersion parameters of $-100\leq v_\mathrm{stellar}\leq100$~km/s and $1\leq \sigma_\mathrm{stellar}\leq300$~km/s, respectively.
This setup is used for 2,038 galaxies (78.2\%). 
Most galaxies with secure redshifts at $z\le1.50$ can achieve this lower $S/N=4$ threshold. 
However, there are 15 sources (0.6\%) that do not reach $S/N=4$ or fail to fit with \gist\ and resolved data products for these cases are not generated.

For emission line fitting, we use a customised version of \textsc{pyGandALF} with two additional features: (1) a jittered start that improves the reliability of fits in low $S/N$ galaxy outskirts where the default \gist\ fitting method often fails to converge (resulting in no output) and (2) the errors on fluxes, velocities, and velocity dispersion for the emission lines are estimated using a Monte Carlo (MC) approach. 
For the jittered start, we run the initial fit three times, once with the initial kinematic guess, and two more times with a $\pm$3 pixel offset. The `best fit' is taken as the fit from these three that returns the minimum chi-squared.
For the MC approach, we take the best-fit spectrum and add randomly shuffled  error-normalised residuals, which are then rescaled by the (un-shuffled) noise array. The parameter uncertainties are taken as the standard deviation among 50 realisations. 
The emission line velocities are restricted to be within $-200\leq v_\mathrm{gas}\leq200$~km/s relative to the stellar continuum velocity of the nearest Voronoi bin, and the velocity dispersions are restricted to be in the range of $1\leq \sigma_\mathrm{gas}\leq300$~km/s. The same gas kinematic restrictions on neighbouring Voronoi bins are adopted regardless of the Voronoi binning thresholds discussed above. 

Similarly to the resolved stellar continuum fitting, we adopt two thresholds for the 1D spectra fits depending on the quality of the data inferred from the spatially-resolved fitting. For the 554 galaxies where two or more Voronoi bins with continuum $S/N\geq10$ were possible, we restrict the parameter range for stellar velocities and velocity dispersion of $-200\leq v_\mathrm{stellar}\leq200$~km/s and $1\leq \sigma_\mathrm{stellar}\leq500$~km/s, respectively. For the 2,038 galaxies with lower continuum $S/N$, we restrict the parameter range for stellar velocities and velocity dispersion of $-100\leq v_\mathrm{stellar}\leq100$~km/s and $1\leq \sigma_\mathrm{stellar}\leq300$~km/s, respectively. The same restrictions on gas kinematics are adopted as used for the resolved fitting ($-200\leq v_\mathrm{gas}\leq200$~km/s and $1\leq \sigma_\mathrm{gas}\leq300$~km/s).

\section{Stellar Mass Comparison}\label{mass_compare}
\begin{figure}
	\includegraphics[width=0.48\textwidth]{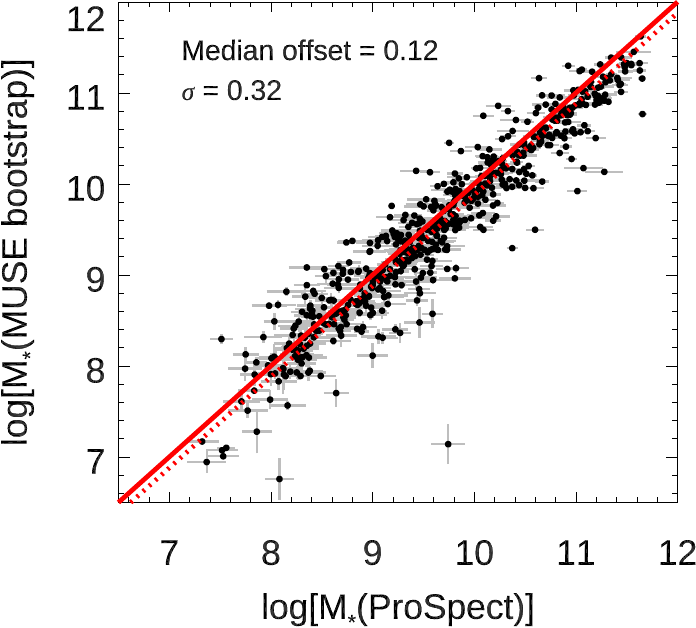}
 \vspace{-5mm}
    \caption{ A comparison between integrated stellar masses from \texttt{ProSpect} SED fitting of optical+near-IR photometry and bootstrapped mass maps from the MUSE data (sum within dilated mask). The red solid line is the 1:1 relation and the dotted red line shows the median offset from the 1:1 relation.
    \label{fig:MAGPI_Mstar_compare_MUSE_ProSpect}}
\end{figure}

As detailed in Section~\ref{sample_overview}, stellar masses based on the MUSE data were bootstrapped to stellar mass maps that use the methodology of \cite{taylor11} derived from Subaru HSC data ($grizy$) for MAGPI fields in G12 and G15 regions. The bootstrapped mass maps are only constructed for sources at $z<0.43$.
A comparison between the bootstrapped stellar masses from the MUSE data and stellar masses from \texttt{ProSpect} based on fits to KiDS+VIKING data ($ugri$+$ZYJHK_s$) is shown in Figure~\ref{fig:MAGPI_Mstar_compare_MUSE_ProSpect}. The MUSE data are deeper and have higher spatial resolution and therefore are preferred over the \texttt{ProSpect} stellar masses. The \texttt{ProSpect} stellar masses are slightly larger with a median offset from a 1:1 relation of log$M_\star$(\texttt{ProSpect})-log$M_\star$(MUSE bootstrap)=0.12~dex. The standard deviation from a 1:1 relation is 0.32~dex.

\section{Tables of Balmer Decrement vs. Effective Radius, Mass Covering Fraction, and Stellar Mass}\label{tables}
In Tables~\ref{tab:bd_gradient_SAMI}, \ref{tab:BDdiff_fc_SAMI}, and \ref{tab:bd_logM_SAMI}  we list the values for the BD-$R/R_e$, BD-$f_c$, and BD-$M_\star$ relations for the SAMI data, respectively. 
In Tables~\ref{tab:bd_gradient_MAGPI}, \ref{tab:BDdiff_fc_MAGPI}, and \ref{tab:bd_logM_MAGPI} we provide the same for the MAGPI data. 

\begin{table}
\caption{Balmer decrement as a function of effective radius (BD gradients) for SAMI galaxies ($0.004 < z< 0.059$) in five mass bins. \label{tab:bd_gradient_SAMI}} 
\begin{center}
\renewcommand{\arraystretch}{1.5} 
\begin{tabular}{cccc}
 \hline \\[-2em]
$\left\langle \log M_\star \right\rangle$ & $N_\mathrm{annuli}$ & $R/R_e$ & \halpha/\hbeta \\
\multirow{7}{*}{8.46} & 598 & 0.16 & $3.07_{-0.41}^{+0.54}$ \\
& 595 & 0.47 & $3.10_{-0.28}^{+0.36}$ \\
& 596 & 0.80 & $3.05_{-0.30}^{+0.34}$ \\
& 598 & 1.14 & $3.02_{-0.40}^{+0.30}$ \\
& 595 & 1.52 & $2.89_{-0.39}^{+0.40}$ \\
& 596 & 2.01 & $2.81_{-0.50}^{+0.41}$ \\
& 597 & 2.85 & $2.58_{-0.53}^{+0.48}$ \\
\hline
\multirow{7}{*}{8.95} & 597 & 0.15 & $3.20_{-0.32}^{+0.41}$ \\
& 596 & 0.48 & $3.13_{-0.23}^{+0.35}$ \\
& 596 & 0.82 & $3.12_{-0.21}^{+0.31}$ \\
& 596 & 1.16 & $3.09_{-0.31}^{+0.27}$ \\
& 596 & 1.53 & $3.01_{-0.35}^{+0.33}$ \\
& 596 & 2.01 & $2.87_{-0.44}^{+0.36}$ \\
& 597 & 2.83 & $2.67_{-0.48}^{+0.49}$ \\
\hline
\multirow{7}{*}{9.29} & 599 & 0.18 & $3.43_{-0.32}^{+0.62}$ \\
& 598 & 0.53 & $3.41_{-0.34}^{+0.49}$ \\
& 599 & 0.88 & $3.32_{-0.29}^{+0.48}$ \\
& 598 & 1.24 & $3.24_{-0.33}^{+0.44}$ \\
& 599 & 1.64 & $3.14_{-0.44}^{+0.45}$ \\
& 598 & 2.14 & $2.97_{-0.50}^{+0.52}$ \\
& 599 & 3.02 & $2.82_{-0.59}^{+0.58}$ \\
\hline
\multirow{7}{*}{9.71} & 596 & 0.16 & $3.80_{-0.42}^{+0.85}$ \\
& 596 & 0.49 & $3.71_{-0.37}^{+0.73}$ \\
& 596 & 0.83 & $3.61_{-0.35}^{+0.68}$ \\
& 596 & 1.18 & $3.54_{-0.35}^{+0.54}$ \\
& 595 & 1.58 & $3.43_{-0.41}^{+0.51}$ \\
& 596 & 2.06 & $3.23_{-0.56}^{+0.47}$ \\
& 596 & 2.92 & $3.00_{-0.59}^{+0.62}$ \\
\hline
\multirow{7}{*}{10.31} & 598 & 0.17 & $4.66_{-0.67}^{+0.92}$ \\
& 598 & 0.50 & $4.36_{-0.49}^{+0.86}$ \\
& 599 & 0.83 & $4.21_{-0.47}^{+0.80}$ \\
& 596 & 1.19 & $4.05_{-0.46}^{+0.76}$ \\
& 598 & 1.59 & $3.86_{-0.51}^{+0.68}$ \\
& 598 & 2.06 & $3.57_{-0.58}^{+0.81}$ \\
& 598 & 2.89 & $3.30_{-0.70}^{+0.78}$ \\
\hline
 \end{tabular}
\end{center}
\end{table}

\begin{table}
\caption{Difference in Balmer decrement as a function of aperture covering fraction (fraction of enclosed stellar mass) for SAMI galaxies ($0.004 < z< 0.059$) in five mass bins. $\Delta \mathrm{BD} = (\mathrm{H}\alpha/\mathrm{H}\beta)_\mathrm{tot}-(\mathrm{H}\alpha/\mathrm{H}\beta)_\mathrm{aper}$ \label{tab:BDdiff_fc_SAMI}} 
\begin{center}
\renewcommand{\arraystretch}{1.5} 
\begin{tabular}{cccc}
 \hline \\[-2em]
$\left\langle \log M_\star \right\rangle$ & $N_\mathrm{aperture}$ & $f_c$ & $\Delta \mathrm{BD}$ \\
\hline
\multirow{7}{*}{8.38} & 259 & 0.03 & $-0.09_{-0.42}^{+0.43}$ \\
& 259 & 0.16 & $-0.07_{-0.25}^{+0.18}$ \\
& 259 & 0.33 & $-0.08_{-0.13}^{+0.13}$ \\
& 259 & 0.50 & $-0.04_{-0.11}^{+0.08}$ \\
& 258 & 0.65 & $-0.03_{-0.08}^{+0.04}$ \\
& 259 & 0.79 & $-0.02_{-0.06}^{+0.02}$ \\
& 259 & 0.93 & $-0.00_{-0.03}^{+0.00}$ \\
\hline
\multirow{7}{*}{8.79} & 259 & 0.04 & $-0.13_{-0.36}^{+0.35}$ \\
& 258 & 0.19 & $-0.09_{-0.23}^{+0.15}$ \\
& 258 & 0.37 & $-0.07_{-0.14}^{+0.10}$ \\
& 259 & 0.54 & $-0.05_{-0.10}^{+0.08}$ \\
& 258 & 0.69 & $-0.03_{-0.06}^{+0.05}$ \\
& 258 & 0.82 & $-0.02_{-0.04}^{+0.02}$ \\
& 259 & 0.93 & $0.00_{-0.02}^{+0.00}$ \\
\hline
\multirow{7}{*}{9.14} & 260 & 0.04 & $-0.11_{-0.31}^{+0.29}$ \\
& 259 & 0.22 & $-0.09_{-0.19}^{+0.14}$ \\
& 259 & 0.41 & $-0.08_{-0.12}^{+0.09}$ \\
& 259 & 0.59 & $-0.06_{-0.08}^{+0.06}$ \\
& 259 & 0.74 & $-0.03_{-0.06}^{+0.03}$ \\
& 259 & 0.86 & $-0.02_{-0.04}^{+0.02}$ \\
& 260 & 0.96 & $-0.00_{-0.03}^{+0.00}$ \\
\hline
\multirow{7}{*}{9.46} & 258 & 0.05 & $-0.15_{-0.36}^{+0.25}$ \\
& 257 & 0.22 & $-0.14_{-0.21}^{+0.13}$ \\
& 257 & 0.41 & $-0.11_{-0.13}^{+0.10}$ \\
& 257 & 0.58 & $-0.07_{-0.08}^{+0.05}$ \\
& 257 & 0.71 & $-0.04_{-0.06}^{+0.03}$ \\
& 257 & 0.83 & $-0.02_{-0.05}^{+0.02}$ \\
& 258 & 0.95 & $-0.01_{-0.03}^{+0.01}$ \\
\hline
\multirow{7}{*}{10.20} & 258 & 0.06 & $-0.40_{-0.57}^{+0.38}$ \\
& 257 & 0.24 & $-0.28_{-0.24}^{+0.20}$ \\
& 257 & 0.42 & $-0.19_{-0.24}^{+0.13}$ \\
& 258 & 0.60 & $-0.13_{-0.13}^{+0.10}$ \\
& 257 & 0.74 & $-0.07_{-0.08}^{+0.06}$ \\
& 257 & 0.86 & $-0.03_{-0.06}^{+0.03}$ \\
& 258 & 0.96 & $-0.01_{-0.04}^{+0.01}$ \\
\hline
 \end{tabular}
\end{center}
\end{table}

\begin{table}
\caption{Integrated aperture Balmer decrements vs. stellar mass for SAMI galaxies ($0.004 < z< 0.059$) in five mass bins. \label{tab:bd_logM_SAMI}} 
\begin{center}
\renewcommand{\arraystretch}{1.5} 
\begin{tabular}{ccc}
 \hline \\[-2em]
 $N_\mathrm{galaxy}$ & $\left\langle \log M_\star \right\rangle$ & \halpha/\hbeta \\
\hline
225 & 8.39 & $2.97_{ -0.21}^{+0.20}$ \\
196 & 8.93 & $3.13_{ -0.21}^{+0.21}$ \\
170 & 9.29 & $3.36_{ -0.30}^{+0.29}$ \\
166 & 9.74 & $3.70_{ -0.37}^{+0.40}$ \\
161 & 10.36 & $4.30_{ -0.53}^{+0.58}$ \\
\hline
 \end{tabular}
\end{center}
\end{table}

\begin{table}
\caption{Balmer decrement as a function of effective radius (BD gradients) for MAGPI galaxies ($0.228 < z< 0.424$) in five mass bins. \label{tab:bd_gradient_MAGPI}} 
\begin{center}
\renewcommand{\arraystretch}{1.5} 
\begin{tabular}{cccc}
 \hline \\[-2em]
$\left\langle \log M_\star \right\rangle$ & $N_\mathrm{annuli}$ & $R/R_e$ & \halpha/\hbeta \\
\hline
\multirow{5}{*}{8.41} &  47 & 0.17 &  $3.89_{-0.75}^{+0.48}$ \\ 
 &  47 & 0.35 &  $3.77_{-0.27}^{+0.65}$ \\ 
 &  47 & 0.62 &  $3.54_{-0.31}^{+0.31}$ \\ 
 &  46 & 0.86 &  $3.48_{-0.68}^{+0.67}$ \\ 
 &  46 & 1.13 &  $3.17_{-0.39}^{+0.74}$ \\ 
 &  46 & 1.44 &  $3.17_{-0.46}^{+0.63}$ \\ 
 &  46 & 1.91 &  $2.88_{-0.48}^{+0.97}$ \\ 
\hline
\multirow{5}{*}{8.98} &  48 & 0.14 &  $3.70_{-0.59}^{+0.72}$ \\ 
 &  47 & 0.38 &  $3.58_{-0.35}^{+0.53}$ \\ 
 &  47 & 0.64 &  $3.46_{-0.23}^{+0.68}$ \\ 
 &  47 & 0.91 &  $3.39_{-0.38}^{+0.44}$ \\ 
 &  47 & 1.22 &  $3.32_{-0.27}^{+0.46}$ \\ 
 &  47 & 1.50 &  $3.16_{-0.28}^{+0.52}$ \\ 
 &  47 & 1.96 &  $3.00_{-0.55}^{+0.92}$ \\ 
\hline
\multirow{5}{*}{9.42} &  46 & 0.14 &  $4.17_{-0.69}^{+1.16}$ \\ 
 &  46 & 0.43 &  $3.93_{-0.46}^{+0.96}$ \\ 
 &  46 & 0.70 &  $3.63_{-0.26}^{+0.64}$ \\ 
 &  46 & 0.96 &  $3.48_{-0.32}^{+0.39}$ \\ 
 &  46 & 1.26 &  $3.29_{-0.39}^{+0.58}$ \\ 
 &  46 & 1.63 &  $3.30_{-0.69}^{+0.19}$ \\ 
 &  45 & 2.12 &  $3.16_{-0.48}^{+0.64}$ \\ 
\hline
\multirow{5}{*}{9.87} &  45 & 0.14 &  $4.45_{-0.61}^{+1.47}$ \\ 
 &  45 & 0.47 &  $4.42_{-0.57}^{+0.98}$ \\ 
 &  45 & 0.75 &  $4.09_{-0.50}^{+0.81}$ \\ 
 &  45 & 1.06 &  $3.76_{-0.34}^{+0.88}$ \\ 
 &  45 & 1.36 &  $3.67_{-0.36}^{+0.67}$ \\ 
 &  45 & 1.68 &  $3.46_{-0.31}^{+1.06}$ \\ 
 &  44 & 2.35 &  $3.28_{-0.34}^{+1.08}$ \\ 
\hline
\multirow{5}{*}{10.50} &  48 & 0.18 &  $5.50_{-0.95}^{+1.08}$ \\ 
 &  48 & 0.52 &  $4.93_{-0.66}^{+1.54}$ \\ 
 &  48 & 0.84 &  $4.69_{-0.64}^{+1.09}$ \\ 
 &  48 & 1.19 &  $4.29_{-0.31}^{+0.51}$ \\ 
 &  48 & 1.54 &  $3.95_{-0.37}^{+0.31}$ \\ 
 &  48 & 1.91 &  $3.63_{-0.28}^{+0.79}$ \\ 
 &  47 & 2.38 &  $3.40_{-0.26}^{+0.69}$ \\ 
\hline
 \end{tabular}
\end{center}
\end{table}

\begin{table}
\caption{Difference in Balmer decrement as a function of aperture covering fraction (fraction of enclosed stellar mass) for MAGPI galaxies ($0.228 < z< 0.424$) in five mass bins. $\Delta \mathrm{BD} = (\mathrm{H}\alpha/\mathrm{H}\beta)_\mathrm{tot}-(\mathrm{H}\alpha/\mathrm{H}\beta)_\mathrm{aper}$ \label{tab:BDdiff_fc_MAGPI}} 
\begin{center}
\renewcommand{\arraystretch}{1.5} 
\begin{tabular}{cccc}
 \hline \\[-2em]
$\left\langle \log M_\star \right\rangle$ & $N_\mathrm{aperture}$ & $f_c$ & $\Delta \mathrm{BD}$ \\
\hline
\multirow{10}{*}{8.41} &  47 & 0.17 &  $-0.39_{-0.39}^{+0.58}$ \\ 
 &  47 & 0.34 &  $-0.29_{-0.36}^{+0.37}$ \\ 
 &  47 & 0.59 &  $-0.27_{-0.19}^{+0.29}$ \\ 
 &  46 & 0.79 &  $-0.06_{-0.21}^{+0.13}$ \\ 
 &  46 & 0.91 &  $-0.02_{-0.14}^{+0.03}$ \\ 
 &  46 & 0.98 &  $-0.01_{-0.11}^{+0.01}$ \\ 
 &  46 & 1.00 &  $ 0.00_{-0.08}^{+0.02}$ \\ 
\hline
\multirow{10}{*}{8.98} &  48 & 0.13 &  $-0.21_{-0.30}^{+0.43}$ \\ 
 &  47 & 0.31 &  $-0.14_{-0.24}^{+0.18}$ \\ 
 &  47 & 0.55 &  $-0.11_{-0.18}^{+0.17}$ \\ 
 &  47 & 0.73 &  $-0.07_{-0.12}^{+0.11}$ \\ 
 &  47 & 0.88 &  $-0.05_{-0.08}^{+0.10}$ \\ 
 &  47 & 0.96 &  $-0.01_{-0.04}^{+0.01}$ \\ 
 &  47 & 1.00 &  $ 0.00_{-0.01}^{+0.01}$ \\ 
\hline
\multirow{10}{*}{9.42} &  46 & 0.13 &  $-0.47_{-0.52}^{+0.46}$ \\ 
 &  46 & 0.36 &  $-0.40_{-0.31}^{+0.32}$ \\ 
 &  46 & 0.58 &  $-0.22_{-0.22}^{+0.18}$ \\ 
 &  46 & 0.74 &  $-0.10_{-0.22}^{+0.10}$ \\ 
 &  46 & 0.87 &  $-0.06_{-0.10}^{+0.06}$ \\ 
 &  46 & 0.95 &  $-0.02_{-0.03}^{+0.02}$ \\ 
 &  45 & 0.99 &  $-0.00_{-0.01}^{+0.01}$ \\ 
\hline
\multirow{10}{*}{9.87} &  45 & 0.10 &  $-0.55_{-0.74}^{+0.39}$ \\ 
 &  45 & 0.35 &  $-0.38_{-0.35}^{+0.32}$ \\ 
 &  45 & 0.57 &  $-0.24_{-0.24}^{+0.21}$ \\ 
 &  45 & 0.74 &  $-0.10_{-0.20}^{+0.12}$ \\ 
 &  45 & 0.86 &  $-0.04_{-0.14}^{+0.05}$ \\ 
 &  45 & 0.94 &  $-0.02_{-0.06}^{+0.01}$ \\ 
 &  44 & 0.99 &  $-0.00_{-0.03}^{+0.00}$ \\ 
\hline
\multirow{10}{*}{10.50} &  48 & 0.14 &  $-0.81_{-0.66}^{+0.41}$ \\ 
 &  48 & 0.46 &  $-0.60_{-0.35}^{+0.36}$ \\ 
 &  48 & 0.69 &  $-0.41_{-0.34}^{+0.26}$ \\ 
 &  48 & 0.83 &  $-0.26_{-0.20}^{+0.15}$ \\ 
 &  48 & 0.92 &  $-0.16_{-0.10}^{+0.09}$ \\ 
 &  48 & 0.97 &  $-0.05_{-0.05}^{+0.04}$ \\ 
 &  47 & 0.99 &  $-0.01_{-0.02}^{+0.01}$ \\ 
\hline
 \end{tabular}
\end{center}
\end{table}

\begin{table}
\caption{Integrated aperture Balmer decrements vs. stellar mass for MAGPI galaxies ($0.228 < z< 0.424$) in five mass bins. \label{tab:bd_logM_MAGPI}} 
\begin{center}
\renewcommand{\arraystretch}{1.5} 
\begin{tabular}{ccc}
 \hline \\[-2em]
 $N_\mathrm{galaxy}$ & $\left\langle \log M_\star \right\rangle$ & \halpha/\hbeta \\
\hline
 69 &  8.42 & $3.51_{-0.35}^{+0.55}$ \\ 
 51 &  8.99 & $3.54_{-0.34}^{+0.39}$ \\ 
 45 &  9.42 & $3.77_{-0.48}^{+0.71}$ \\ 
 35 &  9.85 & $4.22_{-0.54}^{+0.68}$ \\ 
 23 & 10.47 & $4.76_{-0.73}^{+0.91}$ \\ 
\hline
 \end{tabular}
\end{center}
\end{table}


\bsp	
\label{lastpage}
\end{document}